\documentclass[structabstract]{aa} 
\usepackage{CJK}
\usepackage{amsmath}
\usepackage{graphicx}
\usepackage{txfonts}
\usepackage{natbib}
\usepackage{longtable}
\usepackage{lscape}
\usepackage{multirow}
\usepackage{pdfpages}
\usepackage{ulem}
\usepackage[colorlinks=true, allcolors=blue]{hyperref}

\bibpunct{(}{)}{;}{a}{}{,}

\newcommand{\kms}{km\,s$^{-1}$}

\newcommand{\degree}{$^{\circ}$}

\begin{document}
\title{Widespread subsonic turbulence in Ophiuchus North 1 \thanks{Corresponding author: Yan Gong (ygong@mpifr-bonn.mpg.de), Shu Liu (liushu@nao.cas.cn)}}
\author{Yan Gong\inst{1}, Shu Liu\inst{2}, Junzhi Wang\inst{3,4}, Weishan Zhu\inst{5}, Guang-Xing Li\inst{6}, Wenjin Yang\inst{1}, Jixian Sun\inst{7}}

\institute{
Max-Planck-Institut f{\"u}r Radioastronomie, Auf dem H{\"u}gel 69, D-53121 Bonn, Germany
\and 
National Astronomical Observatories, Chinese Academy of Sciences, Beijing 100101, PR China
\and 
Shanghai Astronomical Observatory, Chinese Academy of Sciences, 80 Nandan Road, Shanghai 200030 PR China
\and 
School of Physical Science and Technology, Guangxi University, Nanning 530004, PR China
\and 
School of Physics and Astronomy, Sun Yat-Sen University, Zhuhai campus, No. 2, Daxue Road, Zhuhai, Guangdong 519082, PR China
\and 
South-Western Institute for Astronomy Research, Yunnan University, Kunming, Yunnan 650500, PR China
\and
Purple Mountain Observatory and Key Laboratory of Radio Astronomy, Chinese Academy of Sciences, Nanjing 210034,
PR China
}

\date{Received date ; accepted date}

\abstract
{Supersonic motions are common in molecular clouds. (Sub)sonic turbulence is usually detected toward dense cores and filaments. However, it remains unknown whether (sub)sonic motions at larger scales ($\gtrsim$1~pc) can be present in different environments or not.}
{Located at a distance of about 110 pc, Ophiuchus North 1 (Oph N1) is one of the nearest molecular clouds that allows in-depth investigation of its turbulence properties by large-scale mapping observations of single-dish telescopes.}
{We carried out the $^{12}$CO ($J=1-0$) and C$^{18}$O ($J=1-0$) imaging observations toward Oph N1 with the Purple Mountain Observatory 13.7 m telescope. The observations have an angular resolution of $\sim$55\arcsec (i.e., 0.03~pc).}
{Most of the whole C$^{18}$O emitting regions have Mach numbers of $\lesssim$1, demonstrating the large-scale (sub)sonic turbulence across Oph N1. Based on the polarization measurements, we estimate the magnetic field strength of the plane-of-sky component to be $\gtrsim$9~$\mu$G. We infer that Oph N1 is globally sub-Alfv{\'e}nic, and is supported against gravity mainly by the magnetic field. The steep velocity structure function can be caused by the expansion of the Sh~2-27 H{\scriptsize II} region or the dissipative range of incompressible turbulence.}
{Our observations reveal a surprising case of clouds characterised by widespread subsonic turbulence and steep  size-linewidth relationship. This cloud is magnetized where ion-neutral friction should play an important role.}


\titlerunning{Widespread subsonic turbulence}

\authorrunning{Y. Gong et al.}

\maketitle


\section{Introduction of interstellar turbulence}
Turbulence plays an important role in controlling star formation \citep[e.g.,][]{2004ARA&A..42..211E}. Observations suggest that molecular clouds show supersonic line widths, which is always interpreted as supersonic turbulence \citep[e.g.,][]{1974ARA&A..12..279Z}. Further studies established an empirical size-line width relationship which is well-known as the Larson's relation \citep{1981MNRAS.194..809L}, and the relationship is revised to be $\sigma \propto R^{0.5}$ \citep[$\sigma$ is the velocity dispersion and $R$ is the cloud radius, e.g.,][]{2009ApJ...699.1092H}. This suggests that turbulence energy will decay with decreasing scales. Based on the scale-dependent turbulence energy cascade processes, one would expect to detect (sub)sonic motions at small scales \citep[e.g.,][]{2007ARA&A..45..565M}. Such sonic motions at a typical scale of $\sim$0.1~pc have been detected in the so-called ``coherent cores" where stars are born \citep[e.g.,][]{1998ApJ...504..223G,2010ApJ...712L.116P}. More recent studies have shown that such sonic motions can appear on a larger scale \citep[e.g.,][]{2011A&A...533A..34H,2015A&A...574A.104T,2017A&A...606A.123H,2018A&A...620A..62G,2020ApJ...896..110L} with an extreme case up to 6.5~pc~\citep{2016A&A...587A..97H}, where all the large-scale sonic motions have been found in the dense filamentary structures. It remains unknown whether large-scale (sub)sonic motions can be present in different environments or not. Furthermore, the reason for the formation of large-scale (sub)sonic turbulence remains inconclusive. Large-scale mappings of nearby molecular clouds provide enough information to tackle these questions. 

\section{Oph N1 as a nearby quiescent cloud}
Ophiuchus North 1 (Oph N1) is selected to investigate the turbulence properties in this study. Based on stellar photometric data with Gaia DR2 parallax measurements, the distance to Oph N1 is found to be 109$^{+8}_{-5}\pm 5$ pc \citep{2020A&A...633A..51Z}, and a distance of 110 pc is adopted in this work. Because of its proximity, this cloud can be well resolved even by single-dish telescopes, allowing in-depth investigations of the physics of interstellar turbulence. Oph N1, also known as L260, is a part of Ophiuchus North \citep{1991ApJS...77..647N,2000ApJ...528..817T,2002A&A...385..909T,2012ApJ...754..104H}, where molecular clouds were first mapped in the $J$=$1-0$ transition of $^{12}$C$^{16}$O (hereafter CO) with the 1.2 m Columbia University Sky Survey Telescope \citep[denoted as Complex 4 in][]{1990A&A...231..137D}, revealing a filamentary structure. The molecular cloud complex shows as a dark patch on the H$\alpha$ nebula, indicating that this cloud is situated at the edge of the Sh 2-27 H{\scriptsize II} region around an O9.5V runaway star $\zeta$ Oph \citep{2011AN....332..147H}. $\zeta$ Oph is located at 113 pc \citep[e.g.,][]{2007A&A...474..653V}, strongly suggesting that the H{\scriptsize II} region ionized by $\zeta$ Oph is physically associated with the molecular clouds in Ophiuchus North. 

Figure~\ref{Fig:overview} presents an overview of the Ophiuchus North region where the Stockert-25m 11 cm continuum emission mainly arises from the free-free emission of the extended H{\scriptsize II} region Sh 2-27, the IRAS 100~$\mu$m emission is mainly attributed to cold dust emission, and the WISE 12~$\mu$m emission is dominated by hot dust emission. 
Oph N1 presents a cometary head-tail morphology projected on the periphery of the extended H{\scriptsize II} region Sh 2-27. Oph N1 appears to protrude more deeply into the Sh 2-27 H{\scriptsize II} region than L134E and OphN2. Such protruding morphology is commonly observed in irradiated structures \citep[e.g.,][]{1997A&A...323..931W,2016A&A...591A..40S,2020ApJ...893...91S}, so the morphology further supports the interaction between Oph N1 and Sh 2-27.  

The Ophiuchus North region has also been mapped in CO isotopic lines with the 4 m millimeter telescope of the Nagoya University \citep{1991ApJS...77..647N,2000ApJ...528..817T}. Their observations suggest that star formation in Oph N1 is inactive in contrast to the $\rho$ Oph cloud core and most of $^{12}$C$^{18}$O (hereafter C$^{18}$O) cores are starless. Hence, gas properties are less affected by internal stellar feedback. Two Planck Galactic cold clumps, G008.52+21.84 and G008.67+22.14, were found in Oph N1 \citep{2011A&A...536A..23P,2016A&A...594A..28P}. Higher angular resolution observations of submillimeter continuum emission resolve G008.67+22.14 into two dense cores, L260-SMM1 and L260-SMM2 \citep[e.g.,][]{2002AJ....124.2756V}. L260-SMM1, also named as SCO~01 in \citet{2017A&A...600A..99M}, was found be a Class I protostar associated with outflows \citep{1996A&A...311..858B,2002AJ....124.2756V,2012ApJ...754..104H}, but was reclassified to be a Class II protostar without associated outflows \citep{2017A&A...600A..99M}. Its submilimeter emission is likely dominated by its disk with a mass of $<$0.05~$M_{\odot}$~\citep[e.g.,][]{2002AJ....124.2756V}. L260-SMM2 is a starless core with a mass of $\sim$10~$M_{\odot}$ \citep[e.g.,][]{2002AJ....124.2756V,2012ApJ...754..104H}, which is confirmed by Spitzer nondetection of 70~$\mu$m emission. Both Planck cold clumps are found to show quite narrow line widths of $\sim$0.4~\kms\,in either C$^{18}$O ($J=1-0$) or NH$_{3}$ (2, 2) \citep{2012ApJ...756...76W,1989ApJS...71...89B}, demonstrating low levels of turbulence. However, the large-scale turbulence properties of Oph N1 have been poorly explored. This is why we have carried out dedicated observations to characterise in details the dynamics of this cloud. 

 Our observations are described in Sect.~\ref{Sec:obs}. In Sect.~\ref{Sec:res}, we report our discoveries. The results are discussed in Sect.~\ref{Sec:dis}. Our summary and conclusions are presented in Sect.~\ref{Sec:sum}.

\section{Observations and data reduction}\label{Sec:obs}
\subsection{PMO-13.7 m observations}
Because of their high abundances and low critical densities, the $J=1-0$ transitions of $^{12}$CO, $^{13}$CO, and C$^{18}$O are ideal tracers of the large-scale gas distribution of molecular clouds. Since C$^{18}$O ($J=1-0$) has lower opacities than the other two, it is best suited to investigate the turbulence properties in Oph N1. We carried out simultaneous imaging observations of CO ($J=1-0$) and C$^{18}$O ($J=1-0$) toward Oph N1 with the Purple Mountain Observatory 13.7 m (PMO-13.7 m) telescope during 2021 May 31 -- June 30 (project code: 21A011). We used 3$\times$3 beam sideband separation Superconducting Spectroscopic Array Receiver as the front end and fast Fourier transform spectrometers (FFTSs) as the back ends  \citep{2012ITTST...2..593S}. FFTSs with instantaneous bandwidths of 1 GHz and 200 MHz were used to record the CO ($J=1-0$) and C$^{18}$O ($J=1-0$) signals, respectively. Each FFTS consists of 16,384 channels, and the resulting channel widths are 61.0 kHz and 12.2 kHz for the two FFTS modes. The corresponding velocity spacings are 0.16~\kms\,and 0.03~\kms\,at the rest frequencies of CO ($J=1-0$) and C$^{18}$O ($J=1-0$) for the two FFTS modes, respectively. The on-the-fly method is employed to map Oph N1 at a scanning rate of 50\arcsec~s$^{-1}$ and a dump time of 0.3 seconds \citep{2018AcASn..59....3S}. The mapping was performed alternatively along the right ascension and declination directions in order to reduce striping effects. These observations took about 80 hours in total. 

The standard chopper wheel method was used for calibrations and correcting the atmospheric attenuation \citep{1976ApJS...30..247U}. The antenna temperature, $T^{*}_{\rm A}$, was converted to the main beam temperature, $T_{\rm mb}$, by applying the relation, $T_{\rm mb}=T^{*}_{\rm A}/\eta_{\rm mb}$ , where $\eta_{\rm mb}$ is the main beam efficiency. 
$\eta_{\rm mb}$ is taken to be 49\%\,and 54\%\,for CO ($J=1-0$) and C$^{18}$O ($J=1-0$) according to the telescope's status report\footnote{see Table 2.4.3 in \url{http://www.radioast.nsdc.cn/ztbg/2019-2020ztbgV1.7.pdf}}. The half-power beam widths (HPBWs) are 46\arcsec\,and 50\arcsec\,for CO ($J=1-0$) and C$^{18}$O ($J=1-0$), respectively. Pointing accuracy is within 5\arcsec. The uncertainties of the absolute flux calibration are assumed to be 10\%\,in this work. During the observations, the typical system temperatures were 315--390 K and 141--221~K on a $T^{*}_{\rm A}$ scale for CO ($J=1-0$) and C$^{18}$O ($J=1-0$), respectively. The median rms noise levels are 0.10~K at a channel width of 0.16~\kms\,for CO ($J=1-0$) and 0.28~K at a channel width of 0.03~\kms\,for C$^{18}$O ($J=1-0$). The C$^{18}$O ($J=1-0$) spectra were binned to a channel width of 0.05~\kms\,in order to improve the signal-to-noise ratios of the resulting effective channels. Throughout this paper, velocities are given with respect to the local standard of rest (LSR) while the rest frequencies of CO ($J=1-0$) and C$^{18}$O ($J=1-0$) are set to be 115271.202 MHz and 109782.173 MHz \citep{2005JMoSt.742..215M}.

The spectra were reduced with the GILDAS software \citep{2005sf2a.conf..721P}, and a first-order baseline was subtracted from each spectrum. Raw data were regridded by convolving with a Gaussian kernel of 1/3 HPBWs. After regridding, the effective angular resolutions become 52\arcsec\,and 55\arcsec\,for CO ($J=1-0$) and C$^{18}$O ($J=1-0$), respectively. 

\begin{figure}[!htbp]
\centering
\includegraphics[width = 0.45 \textwidth]{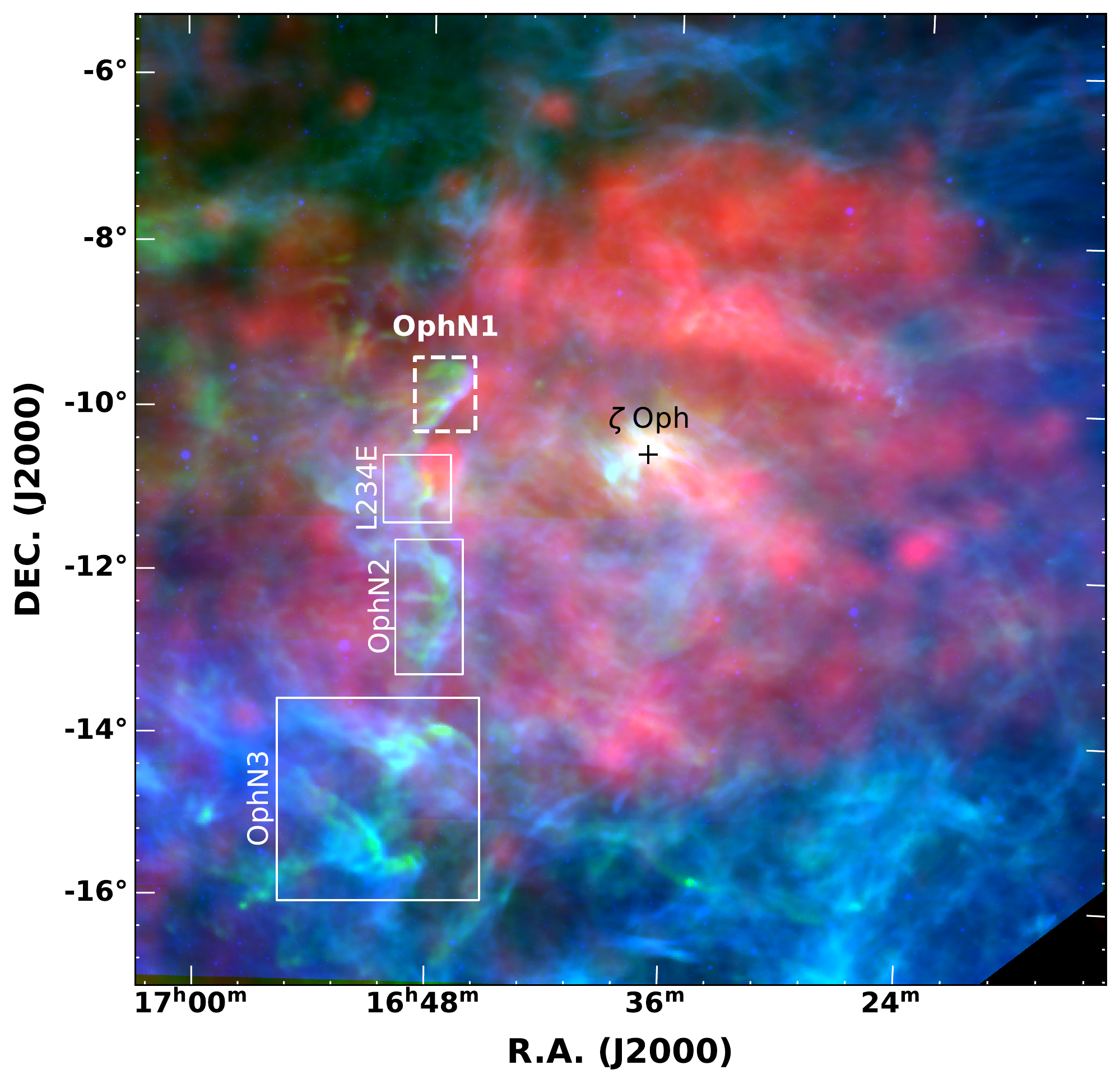}
\caption{{Three-color composite image of the Stockert-25m 11~cm \citep[red;][]{1987MitAG..70..419R}, IRAS 100~$\mu$m \citep[green;][]{2005ApJS..157..302M}, and the WISE 12~$\mu$m \citep[blue;][]{2010AJ....140.1868W} emission toward Ophiuchus North. The Sh 2-27 H{\scriptsize II} region is ionized by $\zeta$ Oph that is indicated by the black cross. The observed region Oph N1 is indicated by the white dashed box, while the other regions are indicated by the white solid boxes.} \label{Fig:overview}}
\end{figure}

\subsection{Archival data}\label{Sec:arc}
 The Planck submillimter continuum polarization data\footnote{The maps can be obtained from the public Planck Legacy Archive \url{http://pla.esac.esa.int/}.} at 353 GHz are employed to study the polarization properties  \citep{2014A&A...571A..11P,2015A&A...576A.105P,2015A&A...576A.104P}. Following previous studies \citep[e.g.,][]{2019A&A...629A..96S}, we smooth the data to an effective angular resolution of 10\arcmin\,to achieve $P/\sigma_{\rm P}>$3, where $P$ is the linear polarization magnitude and $\sigma_{\rm P}$ is the 1$\sigma$ rms level of $P$. In order to be consistent with the IAU convention, the polarization angle of E-vector is calculated with
 \begin{equation}\label{f.pol}
     \Psi_{\rm E} = -0.5 {\rm arctan}(\frac{U}{Q}) \;,
 \end{equation}
The direction of magnetic field, $\Psi_{\rm B}$, is perpendicular to the E-vector, that is $\Psi_{\rm B}=\Psi_{\rm E}-\frac{\pi}{2}$.

We also make use of the near-infrared extinction map to trace the H$_{2}$ column density. The extinction map is derived using the reddening of the light of background stars \citep{2016A&A...585A..78J}. The near-infrared extinction map with an angular resolution of 3\arcmin\,is used in this study. We convert the near-infrared extinction, $A_{\rm J}$, to visual extinction, $A_{\rm V}$, by multiplying by 3.55 \citep{2016A&A...585A..38J}, where the extinction curve, $R_{\rm V} = A_{\rm v}/E(B-V)$, is set to 3.1 \citep{1989ApJ...345..245C}, and $E(B-V)$ is the degree of redenning. We adopt the relationship of \citet{2009MNRAS.400.2050G} to convert $A_{\rm V}$ to H$_{2}$ column density:
\begin{equation}\label{f.av}
    N_{\rm H_{2}} ({\rm cm}^{-2}) = 1.105\times 10^{21} A_{\rm v} ({\rm mag})\;.
\end{equation}

Oph N1 is a high-latitude molecular cloud, and our molecular line observations suggest only one coherent velocity component (see Sect.~\ref{Sec:res}). This supports that there are no additional foreground or background molecular clouds along the line of sight. Hence, the extinction map can well trace the H$_{2}$ column density distribution in Oph N1. 

\section{Results}\label{Sec:res}
\subsection{Molecular distribution}\label{sec.mor}

\begin{figure*}[!htbp]
\centering
\includegraphics[width = 0.95 \textwidth]{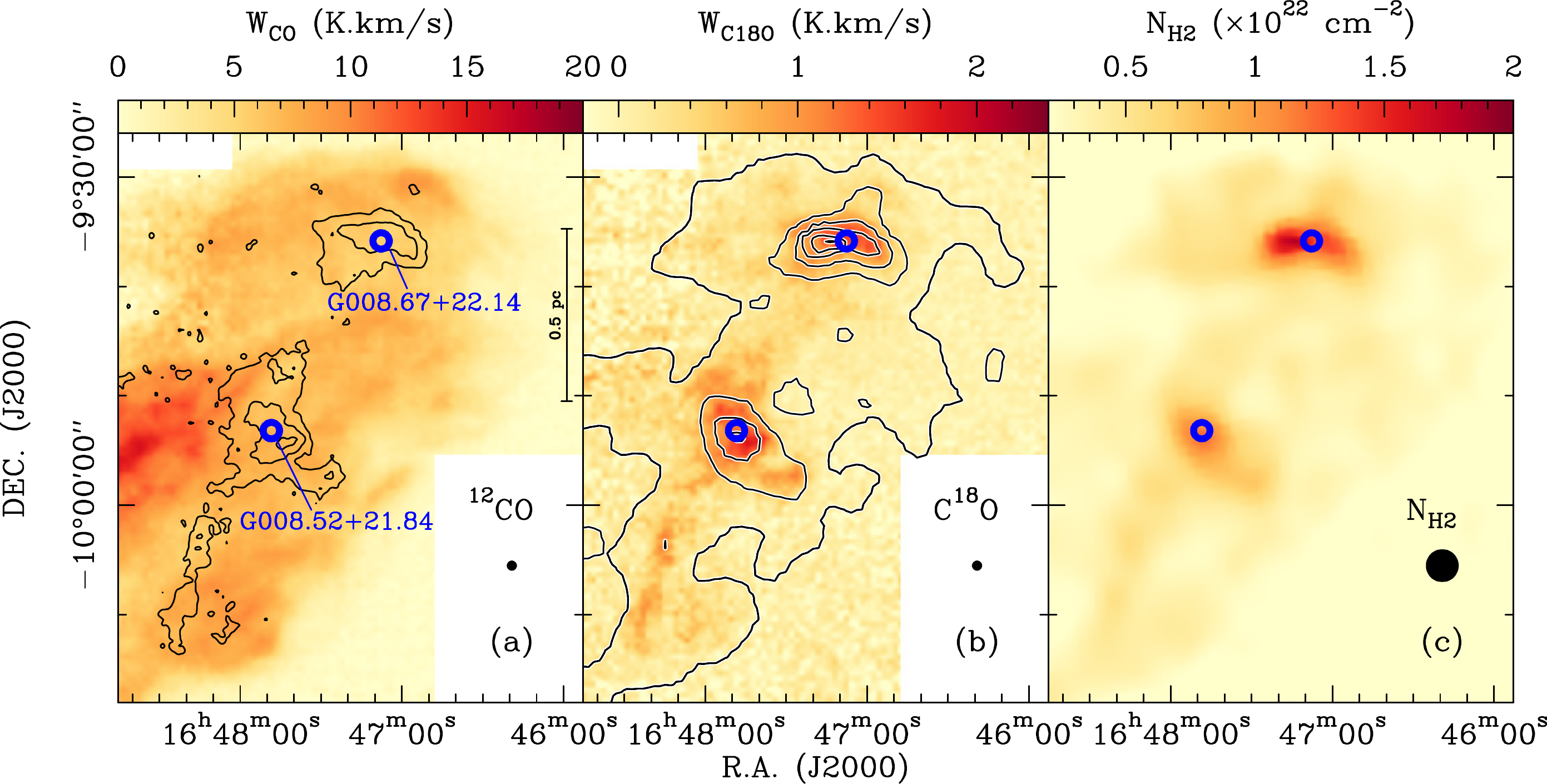}
\caption{(a) CO ($J=1-0$) integrated intensity map, $W_{\rm CO}$, is overlaid with the C$^{18}$O ($J=1-0$) integrated intensity contours. The integrated velocity ranges are [1.3, 6.1]~\kms\,for CO ($J=1-0$) and [2.5, 4.2]~\kms for C$^{18}$O ($J=1-0$). The color bar represents the CO ($J=1-0$) integrated intensity in units of K~\kms, while the contours start at 0.54~K~\kms\,and increase by 0.54~K~\kms. (b) C$^{18}$O ($J=1-0$) integrated intensity map, $W_{\rm C^{18}O}$, is overlaid with the H$_{2}$ column density black contours from the near infrared extinction map \citep{2016A&A...585A..78J}. The color bar represents the C$^{18}$O ($J=1-0$) integrated intensity in units of K~\kms. The black contours start at 3$\times 10^{21}$~cm$^{-2}$ and increase by 3$\times 10^{21}$~cm$^{-2}$. (c) Extinction-based H$_{2}$ column density map. 
In each panel, the beam size is shown in the lower right corner, and the two Planck cold clumps are indicated by the two blue open circles. \label{Fig:co}}
\end{figure*}

Our observed integrated-intensity maps of CO ($J=1-0$) and C$^{18}$O ($J=1-0$), $W_{\rm CO}$ and $W_{\rm C^{18}O}$, are shown in Figs.~\ref{Fig:co}a--\ref{Fig:co}b that provide angular resolutions a factor of $\gtrsim$3 finer than previous CO surveys \citep{1991ApJS...77..647N,2000ApJ...528..817T,2002A&A...385..909T}. The CO ($J=1-0$) integrated-intensity map shows an extended distribution in Fig.~\ref{Fig:co}a. The CO ($J=1-0$) emission above 1~K~\kms\,(3$\sigma$) has a size of about 1.8~pc$\times$1.1~pc with a position angle of $\sim$135\degree, which is similar to the morphology traced by the IRAS 100~$\mu$m emission (see Fig.~\ref{Fig:overview}).
The distribution presents a rift in the northwest, where the C$^{18}$O ($J=1-0$) emission is prominent. This is indicative of the CO ($J=1-0$) self-absorption which can even result in the line ratio [C$^{18}$O/CO] of $>$1 toward G008.52+21.84 (see Appendix~\ref{app.ratio}). 
As shown in Figs.~\ref{Fig:spec}a--\ref{Fig:spec}c, the CO spectra exhibit narrow dips that coincide with the peaks of the corresponding C$^{18}$O ($J=1-0$) spectra, confirming the presence of CO ($J=1-0$) self-absorption. However, the CO ($J=1-0$) self-absorption features are too narrow to be well resolved due to the insufficient spectral resolutions. 

In addition to the difference in the emitting size, the C$^{18}$O ($J=1-0$) morphology is also quite different from that of CO ($J=1-0$). Unlike the extended CO ($J=1-0$) emission, the C$^{18}$O ($J=1-0$) emission exhibits a northwest-southeast filamentary structure with two enhancements at the northwest end and the center where the extinction-based H$_{2}$ column densities exceed 1$\times 10^{22}$~cm$^{-2}$. The two enhancements correspond to the two Planck Galactic cold clumps, G008.52+21.84 and G008.67+22.14 \citep{2011A&A...536A..23P,2016A&A...594A..28P}. The morphology coincides with the infrared extinction map (see Fig.~\ref{Fig:co}b). 



\begin{figure*}[!htbp]
\centering
\includegraphics[width = 0.95 \textwidth]{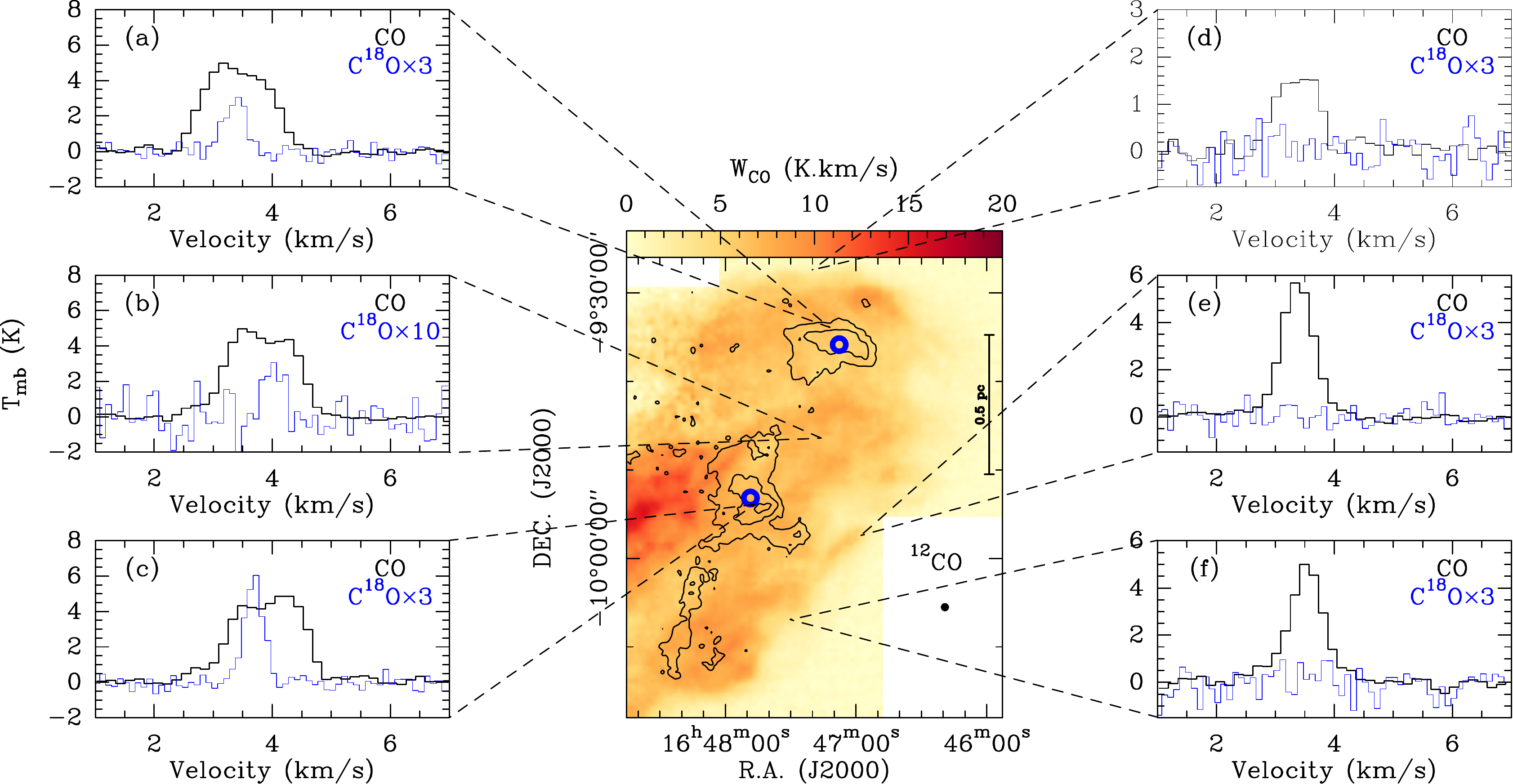}
\caption{CO ($J=1-0$) and C$^{18}$O ($J=1-0$) spectra of the selected positions indicated in the central map that is the same as Fig.~\ref{Fig:co}a. CO ($J=1-0$) and C$^{18}$O ($J=1-0$) are indicated by the black and blue lines, respectively. \label{Fig:spec}}
\end{figure*}

\begin{figure*}[!htbp]
\centering
\includegraphics[width = 0.95 \textwidth]{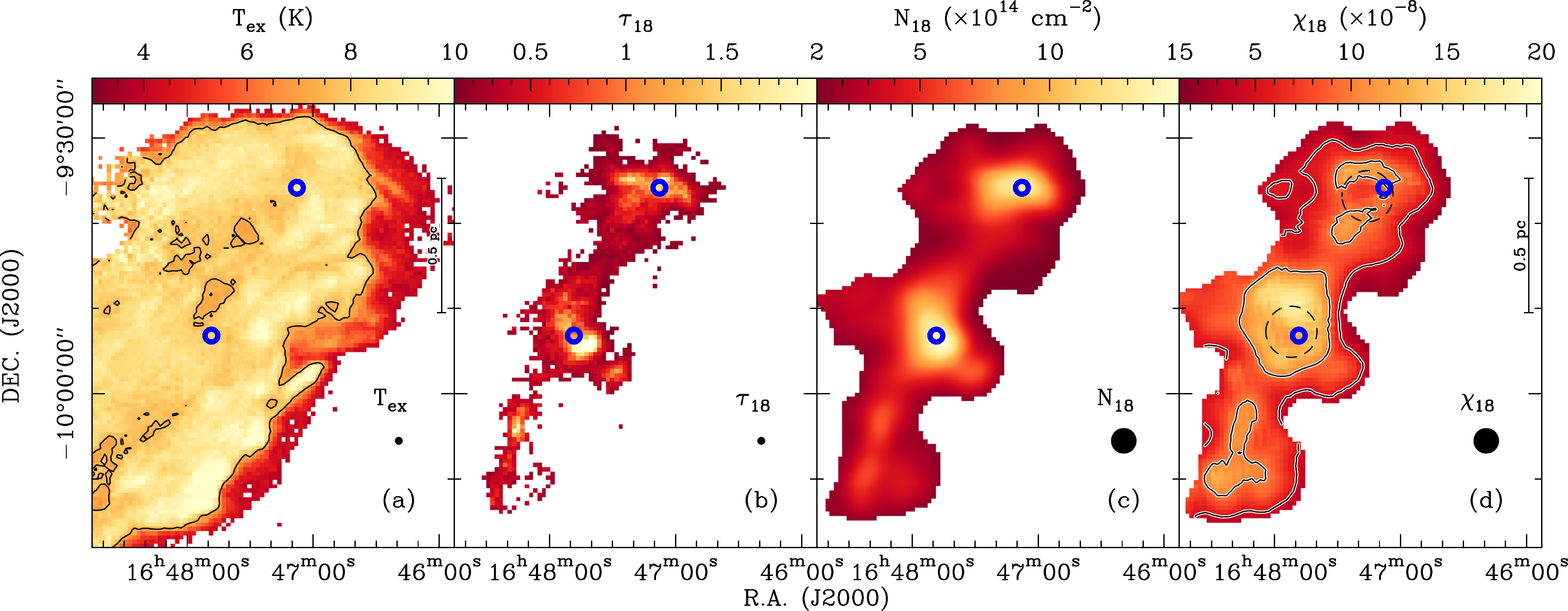}
\caption{(a) Excitation temperature map derived from the CO peak main beam temperature. The color bar represents the excitation temperature in units of K, while the contour represents an excitation temperature of 7.5~K. (b) Peak C$^{18}$O ($J=1-0$) opacity map. (c) C$^{18}$O column density map at an angular resolution of 3\arcmin. (d) C$^{18}$O fractional abundance map. The contours represent the C$^{18}$O fractional abundances of 5$\times 10^{-8}$ and 1$\times 10^{-7}$. The two dashed circles represent the observed CO depletion holes. In each panel, the beam size is shown in the lower right corner, the two Planck cold clumps (G008.67+22.14 and G008.52+21.84) are indicated by the two blue open circles. \label{Fig:exc}}
\end{figure*}

\subsection{Excitation}
Assuming that CO ($J=1-0$) is optically thick (i.e., $\tau_{12}>>$1, where $\tau_{12}$ is the optical depth) and neglecting the beam dilution effects, we are able to derive the excitation temperature, $T_{\rm ex}$, from the CO ($J=1-0$) peak main beam temperature, $T_{\rm p}({\rm CO})$, with the radiative transfer equation \citep[e.g.,][]{2015PASP..127..266M},
\begin{equation}\label{f.rad}
T_{\rm p} = f(J_{\nu}(T_{\rm ex})-J_{\nu}(T_{\rm bg}))(1-{\rm exp}(-\tau)) \;,
\end{equation}

\begin{equation}\label{f.jv}
J_{\nu}(T) = \frac{h\nu/k}{{\rm exp}(h\nu/k)-1}  \;,
\end{equation}
where the beam dilution factor, $f$, is assumed to be unity, ${\rm exp}(-\tau_{12})\sim$0, the background temperature, $T_{\rm bg}$, is set to be 2.73~K \citep{2009ApJ...707..916F},  the Planck constant, $h$, is 6.626$\times 10^{-27}$ erg~s, the Boltzmann constant, $k$, is 1.38$\times 10^{-16}$ erg~K$^{-1}$, and $\nu$ is the rest frequency. 
Oph N1 is seen as a dark patch on the H$\alpha$ nebula of Sh 2-27 \citep{1990A&A...231..137D}, supporting that Oph N1 lies in front of Sh 2-27. Therefore, the continuum emission from the Sh 2-27 H{\scriptsize II} region may contribute to the background temperature term. Based on the 11 cm continuum map \citep{1987MitAG..70..419R}, the brightness temperatures are about 0.4~K. Assuming a typical spectral index for the optically thin free-free continuum emission  (i.e., $T_{\rm B}\sim \nu^{-2.1}$, where $T_{\rm B}$ is the brightness temperature and $\nu$ is the frequency; \citealt{1992ARA&A..30..575C}), we obtain about 0.1~mK at 115 GHz. Hence, the free-free background contribution from the Sh 2-27 H{\scriptsize II} region is negligible in the estimate of the excitation temperature. 

Because the noise distribution is not homogeneous across the observed map, we first estimate the 1$\sigma$ rms noise level of each pixel from emission-free channels, and only take the pixels with CO ($J=1-0$) peak main beam temperatures higher than 5$\sigma$ into account. The derived excitation temperature map is shown in Fig.~\ref{Fig:exc}a. We find that 72\% of pixels have excitation temperatures of 7.5--12.0 K with a median value of 8.5~K (see the distribution within the contour in Fig.~\ref{Fig:exc}a), which is well consistent with the kinetic temperature of 8.8~K derived from previous ammonia observations \citep{1989ApJS...71...89B}. We also note that the excitation temperatures can be underestimated if the beam dilution effects and CO self-absorption become important. 

Molecular gas with excitation temperatures of $>$10~K tends to be clumpy. If the high excitation temperatures are caused by the irradiation from $\zeta$~Oph, this would support a clumpy geometry for photon dominated regions (PDRs) \citep[see][for instance]{2017A&A...598A...2A}. It is also evident that high excitation molecular gas is in the outer regions within the boundary, which is supportive of external heating. The rest 28\% pixels with lower excitation temperatures lie at its outskirt outside the 7.5~K contour boundary in Fig.~\ref{Fig:exc}a. Toward this outskirt, the excitation temperatures drop sharply to $\sim$4 K. This indicates that molecular gas in the outskirt tends to be more diffuse. Furthermore, the optical depths of CO ($J=1-0$) become lower in the outskirt than in the inner regions, possibly violating the optically thick assumption. Therefore, the derived excitation temperatures might be underestimated in the outskirt. Furthermore, the H$_{2}$ number densities become lower, making CO subthermal in the outskirt. One more possibility is that the beam dilution factor becomes significantly lower than unity for CO (1-0) at the very edge of this cloud. 


\subsection{Molecular abundance}
Assuming C$^{18}$O ($J=1-0$) excitation temperature to be the same as CO ($J=1-0$) excitation temperature in Fig.~\ref{Fig:exc}a, we derive the peak C$^{18}$O ($J=1-0$) opacity, $\tau_{18}$, from its peak main beam temperatures above 3$\sigma$ with the radiative transfer equation (see Eq.\ref{f.rad})
and the derived peak C$^{18}$O ($J=1-0$) opacity map is shown in Fig.~\ref{Fig:exc}b. Most of opacities are greater than unity toward the two Planck cold clumps, suggesting that the opacity cannot be neglected to estimate the C$^{18}$O column density in the hubs. In order to improve the signal-to-noise ratios and derive the C$^{18}$O fractional abundance with respect to H$_{2}$, we first smooth the CO ($J=1-0$) and C$^{18}$O ($J=1-0$) data cubes to an angular resolution of 3\arcmin, and recalculate the excitation temperature and C$^{18}$O ($J=1-0$) opacity maps using the same method described above. 

Assuming local thermodynamic equilibrium (LTE), the C$^{18}$O column density, $N_{18}$, can be calculated by using Eqs.~(102) and~(103) in \citet{2015PASP..127..266M},
\begin{equation}\label{f.col}
\begin{split}
N_{18} &=  \frac{3h}{8\pi^{3}\mu^{2}J_{\rm u}} (\frac{kT_{\rm ex}}{h\nu}+1/3){\rm exp}(\frac{E_{\rm u}}{kT_{\rm ex}})[{\rm exp}(\frac{h\nu}{kT_{\rm ex}})-1]^{-1} \\
       & \times \frac{W_{\rm C^{18}O}}{J_{\nu}(T_{\rm ex})-J_{\nu}(T_{\rm bg})} \frac{\tau_{18}}{1-{\rm exp(-\tau_{18})}}\;{\rm cm^{-2}} \;,
\end{split}
\end{equation}
where the dipole moment, $\mu$, is 0.1098~D, $J_{\rm u}$ is the quantum number of the upper level, $W_{\rm C^{18}O}$ is the C$^{18}$O integrated intensity. The resulting map is shown in Fig.~\ref{Fig:exc}c. The derived C$^{18}$O column densities are within the range of (0.7--15.4)$\times 10^{14}$~cm$^{-2}$. The C$^{18}$O fraction abundance with respect to H$_{2}$ is directly determined by the ratio of the C$^{18}$O and H$_{2}$ column densities, and the H$_{2}$ column densities are derived from the extinction. The C$^{18}$O fractional abundance map is shown in Fig.~\ref{Fig:exc}d. The derived abundances are within the range of (0.2--1.7)$\times 10^{-7}$. 


Most emitting regions are consistent with
the typical abundance of (1--2)$\times 10^{-7}$ in nearby dark clouds \citep{1982ApJ...262..590F,1987ApJ...315..621B,1994ApJ...429..694L}. The C$^{18}$O fractional abundance of $<$5$\times 10^{-8}$ is only present in the edge of the cloud (see Fig.~\ref{Fig:exc}d), which is likely caused by the far ultraviolet (FUV) photodissociation of C$^{18}$O \citep[e.g.,][]{1988ApJ...334..771V,2014A&A...564A..68S,2019ApJS..243...25W}. On the other hand, the C$^{18}$O fractional abundances become higher in the two Planck cold clumps, because FUV radiation is well shielded by their high extinctions (i.e., corresponding H$_{2}$ column densities of $\gtrsim 7\times 10^{21}$~cm$^{-2}$).  Ring-like structures are revealed toward the two Planck cold clumps (see the two dashed circles in Fig.~\ref{Fig:exc}d), which can be readily explained by the CO depletion onto dust grains in dense regions \citep[e.g.,][]{2007ARA&A..45..339B}. Both the depletion holes have radii of about 0.1~pc, similar to the typical depletion size in prestellar cores \citep[$\sim$0.1~pc;][]{2007ARA&A..45..339B}. Because of the abundance variations caused by photodissociation and CO depletion, the mass is estimated using the extinction-based H$_{2}$ column density instead. Using the same mask in Fig.~\ref{Fig:exc}a, we derive the total molecular gas mass to be 132~M$_{\odot}$.




\subsection{Velocity field}\label{sec.kin}
\subsubsection{Decomposition}
Because CO ($J=1-0$) is likely optically thick and suffers from self-absorption, CO ($J=1-0$) spectra cannot well trace the velocity field of Oph N1. In contrast, C$^{18}$O ($J=1-0$) has much lower opacities, so its spectra can better trace intrinsic velocity centroids and velocity dispersions of molecular clouds. Therefore, we use C$^{18}$O ($J=1-0$) data to study the kinematics of Oph N1. 

\begin{figure*}[!htbp]
\centering
\includegraphics[height = 0.75 \textwidth]{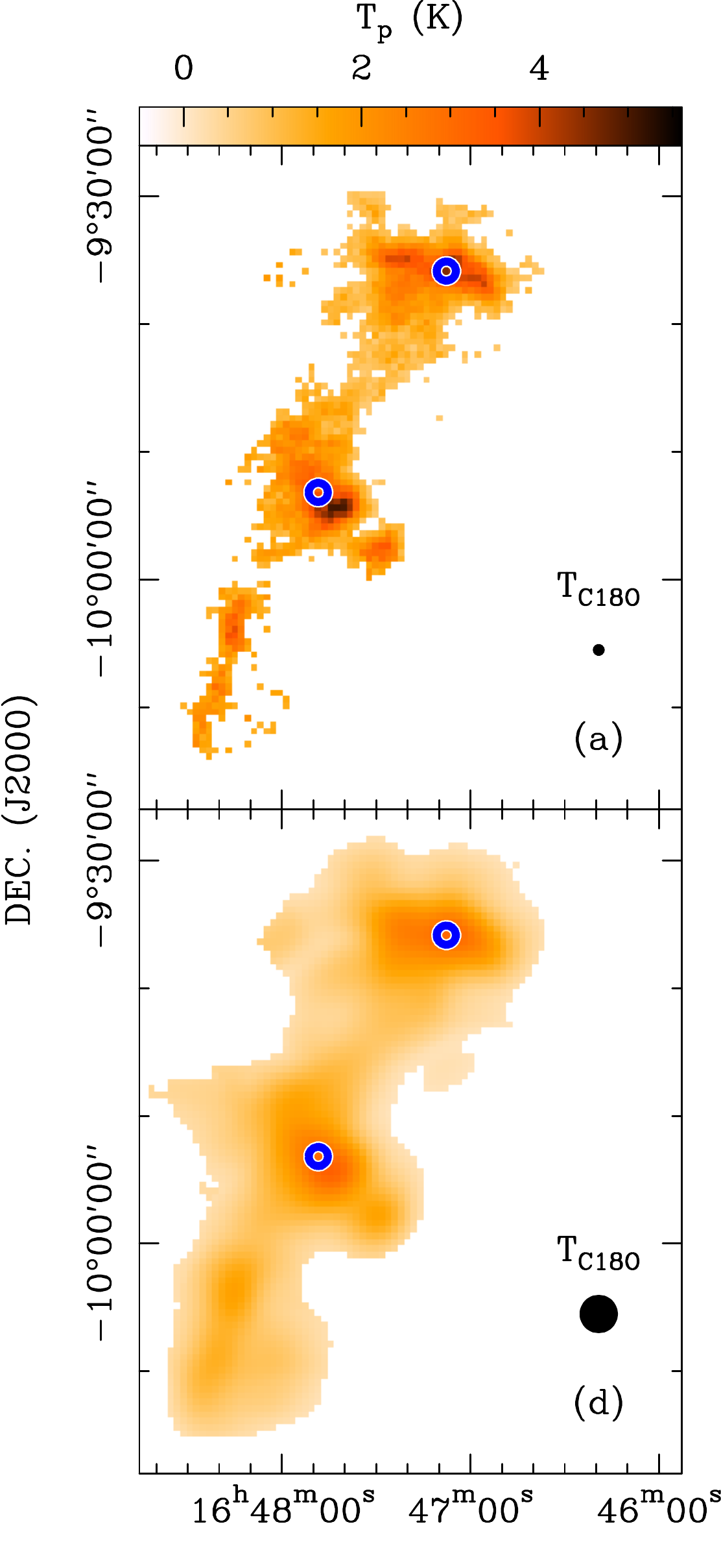}
\includegraphics[height = 0.75 \textwidth]{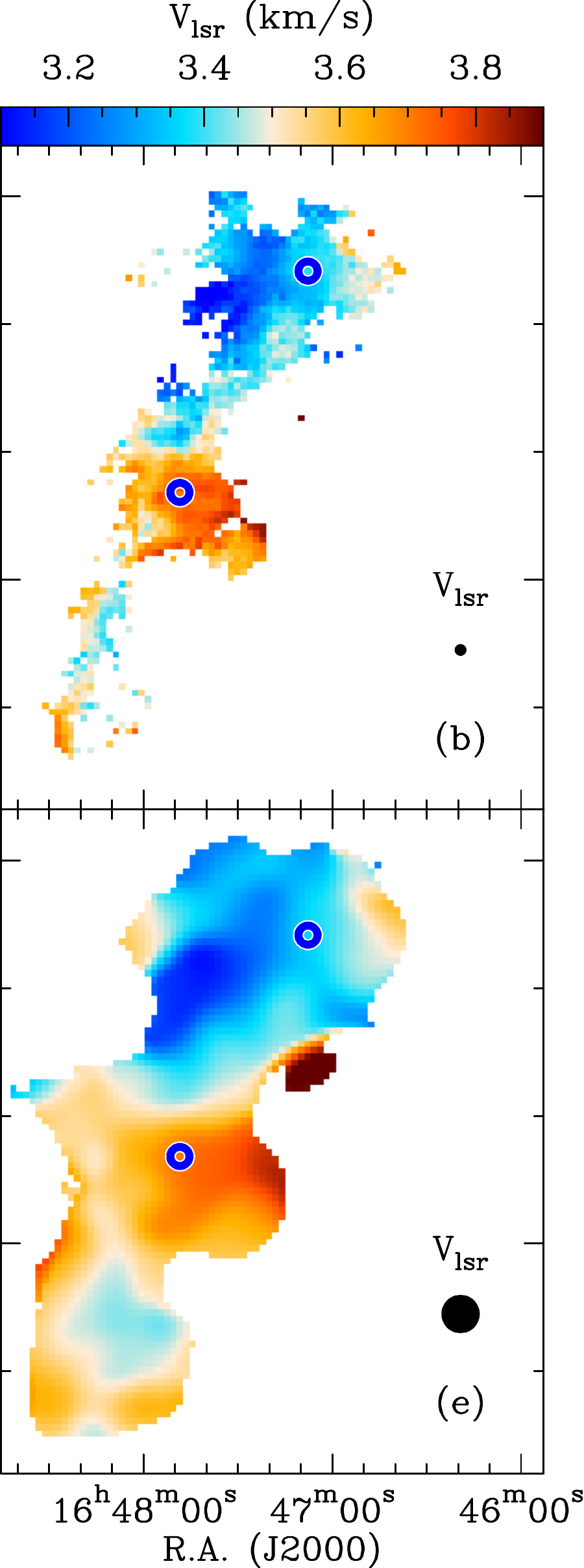}
\includegraphics[height = 0.75 \textwidth]{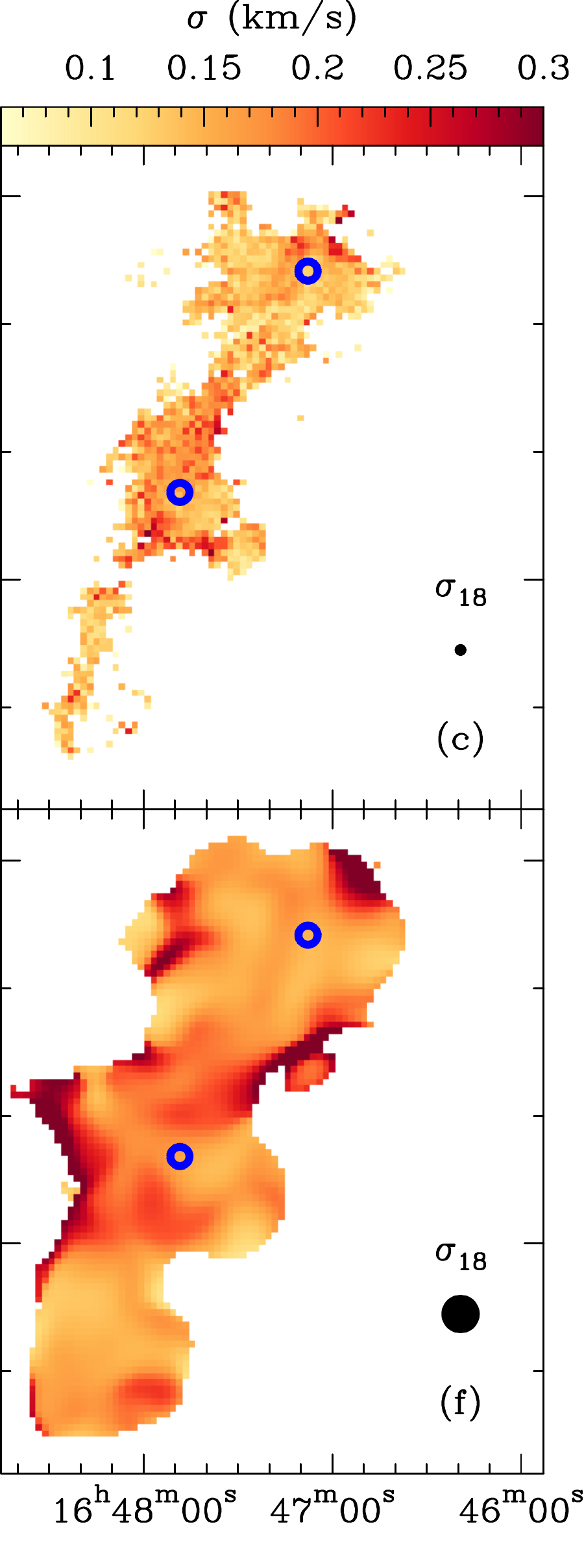}
\caption{Maps of peak intensities (Fig.~\ref{Fig:decom18}a), LSR velocities (Fig.~\ref{Fig:decom18}b), and velocity dispersions (Fig.~\ref{Fig:decom18}c) derived from single-component Gaussian fits to our C$^{18}$O ($J=1-0$) data cube. 
Figures~\ref{Fig:decom18}d--\ref{Fig:decom18}f are similar to Figs.~\ref{Fig:decom18}a--\ref{Fig:decom18}c, but at a smoothed angular resolution of 180\arcsec. 
In each panel, the beam size is shown in the lower right corner, and the two Planck cold clumps (G008.67+22.14 and G008.52+21.84) are indicated by the two open blue circles. 
\label{Fig:decom18}}
\end{figure*}


A Gaussian decomposition allows us to study the kinematic properties of C$^{18}$O ($J=1-0$) data across Oph N1. We manually examine the data cube which appears to display only one single Gaussian component for the whole region. Therefore, we assume a single Gaussian component to decompose the C$^{18}$O ($J=1-0$) spectra with peak intensities higher than 3$\sigma$. Figures~\ref{Fig:decom18}a--\ref{Fig:decom18}c present the distribution of the peak intensities, LSR velocities, and velocity dispersions. We also perform Gaussian convolution on the C$^{18}$O data cube to improve the signal-to-noise ratios at the expense of angular resolution. The C$^{18}$O data cube is smoothed to a beam size of 180\arcsec. We carry out the same decomposition to the smoothed data cube, and the results are shown in Figs.~\ref{Fig:decom18}d--\ref{Fig:decom18}f. The distributions of the peak intensities, LSR velocities, and velocity dispersions are similar to Figs.~\ref{Fig:decom18}a--\ref{Fig:decom18}c, except that the velocity dispersions are slightly higher than the values derived from the data cube with a beam size of 55\arcsec. This is because the velocity gradients tend to be larger in a larger beam. 


\subsubsection{Local velocity gradient}
\begin{figure*}[!htbp]
\centering
\includegraphics[width = 0.95 \textwidth]{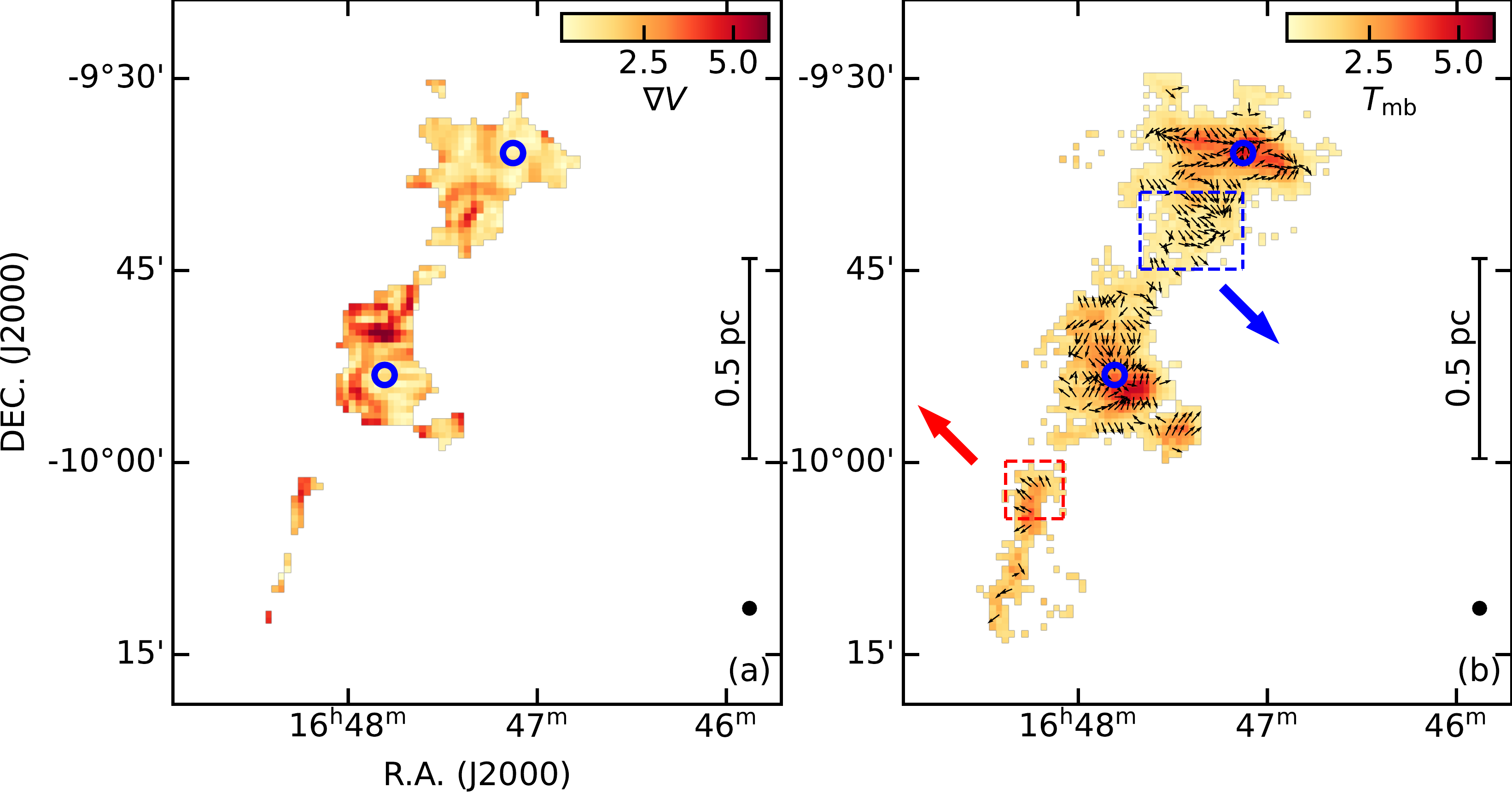}
\includegraphics[width = 0.95 \textwidth]{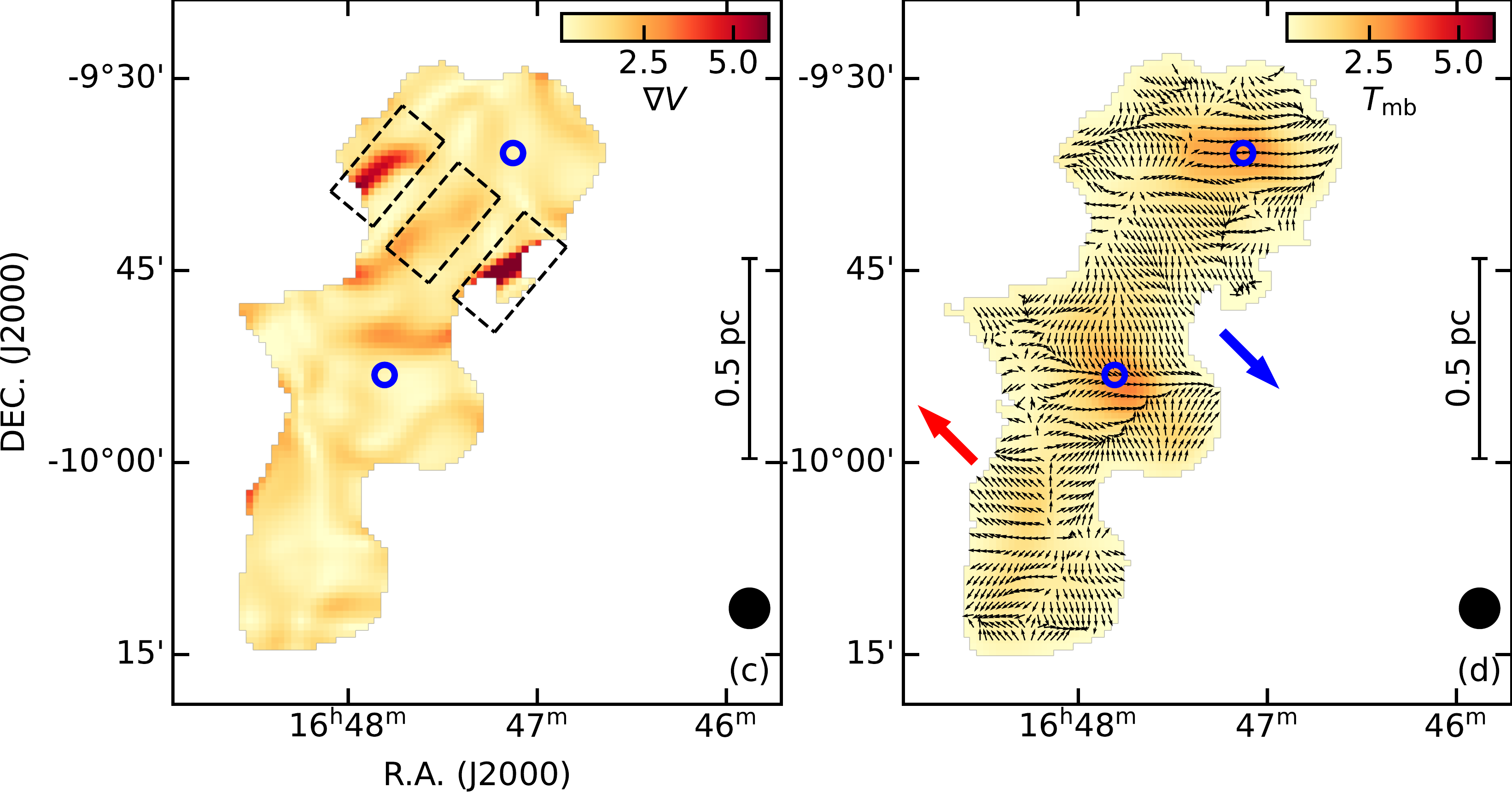}
\caption{(a) Local velocity gradient magnitude map derived from Fig.~\ref{Fig:decom18}b. (b) C$^{18}$O ($J=1-0$) peak intensity map is overlaid with the normalized velocity gradient vectors. The blue and red arrows represent the average directions of the local velocity gradients in sub-regions indicated by the blue and red dashed boxes, respectively. Figures~\ref{Fig:vg}c and \ref{Fig:vg}d are similar to Fig.~\ref{Fig:vg}a and \ref{Fig:vg}b, but derived from Fig.~\ref{Fig:decom18}e. In Fig.~\ref{Fig:vg}c, the three regions showing high velocity gradients are indicated by the three black dashed boxes. In each panel, the beam size is shown in the lower right corner, the two Planck cold clumps (G008.67+22.14 and G008.52+21.84) are indicated by the two open blue circles. \label{Fig:vg}}
\end{figure*}

In the velocity-centroid maps (see Figs.~\ref{Fig:decom18}b and \ref{Fig:decom18}e), velocity centroids range from 3.1~\kms\,to 3.95~\kms. The narrow and continuous distribution suggests that Oph N1 is velocity-coherent. Furthermore, different velocity gradients are evident in different regions. In order to better visualize the local velocity gradients, $\nabla V$, we use the definition of the local velocity gradients by \citet{1993ApJ...406..528G}:
\begin{equation}\label{f.grad}
     \varv_{\rm lsr} = \varv_{0} +a\Delta \alpha +b\Delta \delta\;,
\end{equation}
where $\varv_{\rm lsr}$ is the observed LSR velocity centroid, $\varv_{0}$ is the systemic LSR velocity, $\Delta \alpha$ and $\Delta \delta$ are the offsets in right ascension and declination, and $a$ and $b$ are the components of $\nabla V$ along the directions of right ascension and declination.
The magnitude of $\nabla V$ is $|\nabla V| =\sqrt{a^{2}+b^{2}}$, and the position angle is $\theta_{\rm pa}={\rm arctan}(a/b)$ where $\theta_{\rm pa}$ increases counter-clockwise with respect to the north. 
Following \citet{2021A&A...646A.170G}, we used the Levenberg-Marquardt algorithm to fit this function toward each square block with adjacent 3$\times$3 pixels toward Fig.~\ref{Fig:decom18}b and \ref{Fig:decom18}e. The distribution of the local velocity gradients, $\nabla \varv$, is shown in Fig.~\ref{Fig:vg}.

Figures~\ref{Fig:vg}a and \ref{Fig:vg}b mainly trace the $\nabla \varv$ distributions around the two Planck cold clumps, while Figures~\ref{Fig:vg}c and \ref{Fig:vg}d can better trace $\nabla \varv$ in more extended regions. In Fig.~\ref{Fig:vg}a, high $|\nabla \varv|$ of $\gtrsim$5~\kms~pc$^{-1}$ are found in a shell that has a radius of $\sim$0.1~pc centered at G008.52+21.84, but $|\nabla \varv |$ decreases toward the center. Figure~\ref{Fig:vg}b exhibits a converging morphology toward G008.52+21.84, indicating core accretion from ambient clouds. Th trend of having a decreasing $|\nabla \varv|$ towards the center of the cores has also been detected in other studies \citep{2020ApJ...891...84C,2021A&A...646A.170G}. Because the decreasing $\nabla \varv$ trend is opposite to the prediction by the gravity-driven accretion model \citep[e.g.,][]{2009ApJ...704.1735H}, \citet{2020ApJ...891...84C} proposed that the accretion can be damped by the high-density materials. The other possibility is at least partially due to the geometric effect \citep{2021A&A...646A.170G}. For a spherically collapsing core, the observed LSR velocity should be constant across the core, because the observed LSR velocity is actually density-weighted 3D velocity average over the line of sight. On the other hand, the velocity gradient should be more significant for other geometry like sheets (see Fig.~1 in \citealt{2019A&A...623A..16S} for example). Hence, the decreasing $\nabla \varv$ toward the center of G008.52+21.84 could be due to the fact that the structure becomes more symmetric toward dense regions. In Fig.~\ref{Fig:vg}b, the $\nabla \varv$ converging morphology toward G008.67+22.14 is not as evident as G008.52+21.84, which is likely dominated by the large-scale east-west velocity gradient (see Fig.~\ref{Fig:vg}d).

$\nabla \varv$ in the red and blue dashed boxes appear to be dominated by a northeast-southwest velocity gradient which is almost perpendicular to the filament's long axis and the plane-of-the-sky magnetic field. Such transverse velocity gradients can be caused by filament rotation \citep[e.g.,][]{2020A&A...642A..76Z,2022MNRAS.509.5237S}. However, $\nabla \varv$ in the red and blue dashed boxes show opposite directions, ruling out the filament's rigid rotation. Instead, such a $\nabla \varv$ morphology can be explained by differential rotation or shear motions that are expected in turbulence vorticity \citep[e.g.,][]{2000MNRAS.311...85F,lesieur2008turbulence,2018MNRAS.473.3454B}.

Comparing the results at different angular resolutions, we find that the magnitude of $\nabla \varv$ in Fig.~\ref{Fig:vg}c is lower than Fig.~\ref{Fig:vg}a due to its lower spatial resolution, while the opposite directions of $\nabla \varv$ indicated by the arrows in Fig.~\ref{Fig:vg}a is largely retained in Fig.~\ref{Fig:vg}d. Another interesting feature is that three regions show higher $\nabla \varv$ magnitudes than ambient regions and the three regions are nearly parallel to each other (see the black dashed boxes in Fig.~\ref{Fig:vg}c). The high $\nabla \varv$ magnitudes contribute to the velocity dispersion measured within a beam, leading to the higher velocity dispersions of the corresponding pixels in Fig.~\ref{Fig:decom18}f than in Fig.~\ref{Fig:decom18}c. Such sharp variations have been detected and interpreted as velocity shear by previous studies \citep{2009A&A...500L..29H,2009A&A...507..355F}. Hence, this implies that the observed velocity field is partially regulated by shear motions. 

On the other hand, simulations have shown that turbulence can also cause small-scale fluctuations on the measured velocity centroids  \citep{2022MNRAS.509.5237S}, which should affect the patterns in our $\nabla \varv$ maps (Fig.~\ref{Fig:vg}). However, it is difficult to quantify the contributions of the ordered motions and turbulent motions in Fig.~\ref{Fig:vg} based on our current observations. Nevertheless, large-scale velocity gradients should be much less affected than small-scale velocity gradients by turbulence. 


\subsubsection{Widespread subsonic turbulence}\label{sec.subsonic}
\begin{figure}[!htbp]
\centering
\includegraphics[width = 0.45\textwidth]{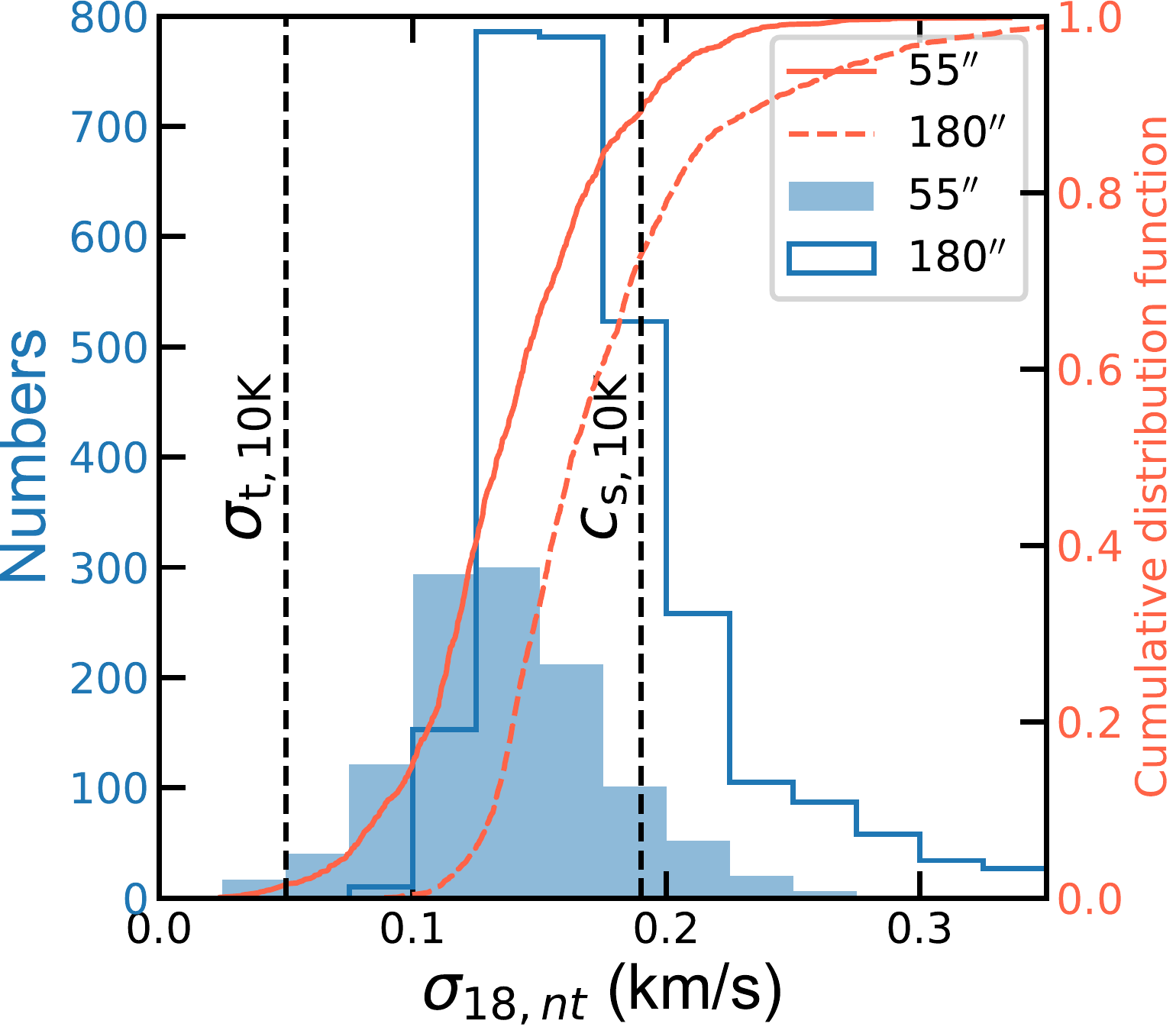}
\caption{{Histogram and cumulative distribution of the nonthermal velocity dispersion derived from C$^{18}$O ($J=1-0$) at the angular resolutions of 55\arcsec\,and 180\arcsec, respectively. The vertical dashed lines represents the sonic speed and the thermal velocity dispersion of C$^{18}$O at a kinetic temperature of 10~K. This figure is suggestive of widespread subsonic turbulence.}\label{Fig:histv}}
\end{figure}

In Fig.~\ref{Fig:decom18}c, we find that all the derived velocity dispersions are lower than 0.25~\kms. The fitting errors in the derived velocity dispersion range from 0.004~\kms to 0.04~\kms. Hence, the derived velocity dispersions should be robust. The observed velocity dispersions come from thermal and nonthermal motions. The non-thermal velocity dispersion, $\sigma_{\rm nt}$, can be estimated by subtracting the thermal velocity dispersion, $\sigma_{\rm t}$, from the observed total velocity dispersion\footnote{In this work, $\sigma_{\rm nt}$, $\sigma_{\rm t}$, and $\sigma_{\rm obs}$ refer to one-dimensional velocity dispersions rather than three-dimensional velocity dispersions.}, $\sigma_{\rm obs}$,
\begin{equation}\label{f.dis}
\sigma_{\rm obs} = \sqrt{\sigma_{\rm nt}^{2} +\sigma_{\rm t}^{2}}\;,
\end{equation}
where $\sigma_{\rm t} = \sqrt{\frac{{\rm k}T}{m_{\rm i}}}$, k is the Boltzmann constant, and $m_{\rm i}$ is the mass weight that is 30 for C$^{18}$O. Here, we adopt a kinetic temperature of 10~K for the fiducial case, which results in $\sigma_{\rm t}$=0.05~\kms\,for C$^{18}$O and a sound speed, $c_{\rm s}$, of 0.19~\kms\, where $m_{\rm i}$=2.37 \citep{2008A&A...487..993K}. The observed Mach number, $\mathcal{M}$, is determined by $\mathcal{M} = \sigma_{\rm nt}/c_{\rm s}$. As shown in Fig.~\ref{Fig:histv}, we find that more than 85\% and 70\%\,of the fitted pixels that have $\mathcal{M} < 1$ at the angular resolutions of 55\arcsec\,and 180\arcsec, respectively. 

\begin{figure}[!htbp]
\centering
\includegraphics[width = 0.45\textwidth]{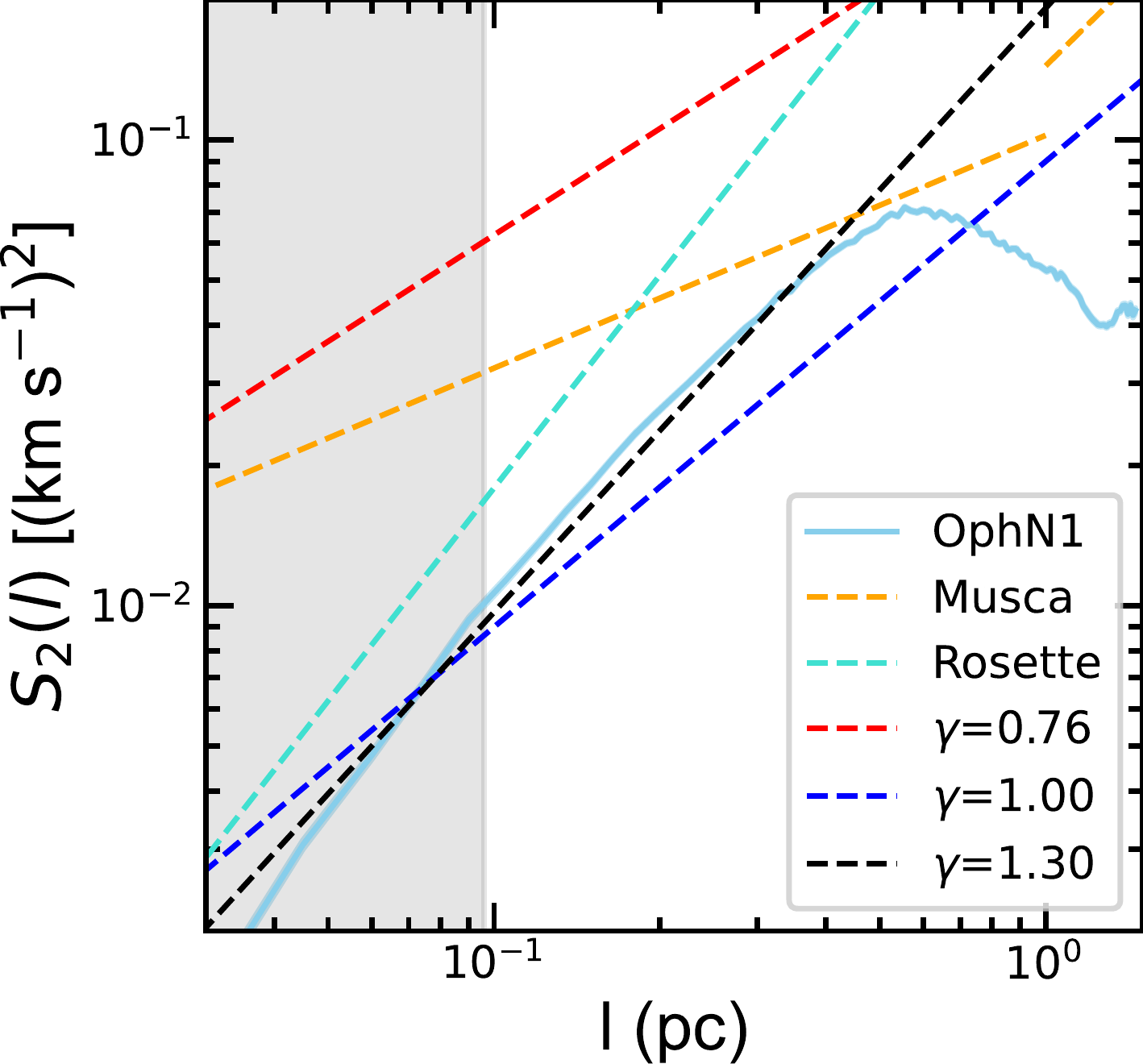}
\caption{{Second-order velocity structure function of Oph N1 derived from Fig.~\ref{Fig:decom18}b as a function of the spatial lag, $l$. 
The gray-shaded region is the spatial-resolution limit which corresponds to 55\arcsec. 
The power-law fitting result is indicated by the black dashed line. The orange, red, and blue dashed lines represent the observed relation in Musca \citep{2016A&A...587A..97H} and Rossete \citep{2006ApJ...643..956H}, the classic Larson relation \citep[$\gamma$=0.76;][]{1981MNRAS.194..809L}, and the revised relation, $\sigma \propto R^{0.5}$ \citep[$\gamma$=1;][]{2009ApJ...699.1092H}, respectively.}\label{Fig:vsf}}
\end{figure}

The observed $\sigma_{\rm nt}$ can also be higher than the intrinsic turbulence velocity dispersion, because the variation of LSR velocity centroids within the beam can partially arise from the ordered motions (e.g., rotation, shear motions) rather than pure turbulence (see also \citealt{2022MNRAS.509.5237S}). Making use of the derived local velocity gradients, we could roughly estimate the contributions to the observed $\sigma_{\rm nt}$. The observed $|\nabla \varv|$ vary from 1~\kms~pc$^{-1}$\,to 10~\kms~pc$^{-1}$ in Figs.~\ref{Fig:vg}a and \ref{Fig:vg}c, suggesting a plane-of-sky contribution of 0.1~\kms\,to 0.3~\kms\,to observed line widths within the beam sizes of 55\arcsec\,and 180\arcsec, respectively. If we assume that the observed $|\nabla \varv|$ is not attributed to turbulence, we can subtract the contributions of ordered motions from the velocity dispersions for each pixel. This correction leads to more than 92\% and 82\%\,of the fitted pixels showing the subsonic level of turbulence at the angular resolutions of 55\arcsec\,and 180\arcsec, respectively, which further reinforces the widespread subsonic turbulence in Oph N1.

Such subsonic turbulence has also been reported in L1517 \citep{2011A&A...533A..34H}, but the prevalence of subsonic motions is up to about 0.5~pc. In contrast, our observations demonstrate large-scale subsonic motions up to a scale of $\gtrsim$1.5~pc, larger than the previous study. Based on the classic size-line width relations \citep[e.g.,][]{1981MNRAS.194..809L,2009ApJ...699.1092H}, sonic motions are expected on scales of $\lesssim$0.3~pc. Therefore, the turbulence properties of Oph N1 apparently violate the classic size-line width scaling relationship.

We also investigate the CO ($J=1-0$) spectra in the outskirts of Oph N1 (Fig.~\ref{Fig:spec}d--\ref{Fig:spec}e). These spectra are found to have line widths of 0.64--0.81~\kms. Removing the broadening effects caused by the channel width and thermal motions, we obtain the nonthermal velocity dispersions of 0.26--0.37~\kms, corresponding to $\mathcal{M}$=1.4--1.8, that is, transonic. Because of the potential opacity broadening effects in CO ($J=1-0$) and local velocity gradients within the beam, the derived $\mathcal{M}$ are upper limits, and the intrinsic motions can be actually more quiescent. Gaussian decomposition to the whole CO ($J=1-0$) data cube confirms the quiescent motions in a larger region (see Appendix~\ref{app.b}). In combination with our C$^{18}$O measurements, our observations indicate that the entire cloud could be fully decoupled from the supersonic environment.





\subsubsection{Steep velocity structure function}\label{sec.vsf}
Velocity structure functions are useful to study the dynamical state of molecular clouds \citep[e.g.,][]{1994ApJ...429..645M,2002A&A...390..307O,2004ApJ...615L..45H,2005ApJ...631..320E,2019A&A...630A..97C,2020NatAs...4.1064H}. Following previous studies \citep[e.g.,][]{2019A&A...630A..97C}, the second-order velocity structure function, $S_{2}$, is a two-point correlation function that quantifies the mean velocity difference:
\begin{equation}
  S_{2} = <\delta \varv^{2}>= <|\varv(x+l) -\varv(x)|^{2}> \approx l^{\gamma}
\end{equation}
where $l$ is the spatial lag between two positions, $x$ and $x+l$, and $\gamma$ is the power-law index.

Because Fig.~\ref{Fig:decom18}e has a higher dynamical range than Fig.~\ref{Fig:decom18}b and the bulk motions are nearly identical at large scales, we make use of Fig.~\ref{Fig:decom18}e to derive the second-order velocity structure function for Oph N1, and the result is shown in Fig.~\ref{Fig:vsf}. The distribution appears to be linear at the spatial lag of $l<$0.5~pc, because the completeness limit is about 0.5~pc in our study. Hence, a linear fit is performed on the data in the logarithmic form in order to derive $\gamma$ in the spatial range of 0.03--0.5~pc. The resulting $\gamma$ is 1.30$\pm$0.03, which is steeper than previous reported values of molecular clouds in the Galactic disk \citep[see Table~1 in][]{2019A&A...630A..97C} and the classical Larson relation \citep[e.g.,][]{1981MNRAS.194..809L,2009ApJ...699.1092H}, but is similar to the reported size-line width relations in the Rosette molecular cloud \citep{2006ApJ...643..956H}, Taurus $^{13}$CO cores  \citep[$\gamma\sim$1.4;][]{2012ApJ...760..147Q}
, the central molecular zone \citep[$\gamma$=1.32$\pm$0.36;][]{2017A&A...603A..89K}, and nearby young stellar object associations measured with the Gaia proper motions \citep[$\gamma\sim$1.3;][]{2021arXiv211011595Z}.  

The derived $\gamma$ cannot be explained by the classic energy cascade that predicts $\gamma$=2/3 \citep{1941DoSSR..30..301K}. \citet{2017A&A...603A..89K} proposed the decay of gas motions to transonic velocities in strong shocks to explain the observed steep size-line width relation. This scenario appears to contradict widespread subsonic motions observed in Oph N1. However, we cannot rule out the possibility that the observed structure is a fossil of shocks, in which kinetic energies have been already dissipated. Based on the comparison of clouds inside and outside the ionization front induced by the Rosette clusters \citep{2006ApJ...643..956H}, the steep index is thought to arise from the expansion of ionized gas. Because Oph N1 is interacting with the Sh~2-27 H{\scriptsize II} region, this scenario is favored to explain our case. On the other hand, the steep index indicates that Oph N1 is dissipating more efficiently than the classic energy cascade. In the frame of incompressible turbulence that is expected in the subsonic regime, the index in the dissipative range is steeper than in the inertial range. Furthermore, filamentary structures are known to be typical structures of turbulence dissipation for low Mach numbers \citep[e.g.,][]{2007ApJ...665L..35D,Zybin_2015,2020FlDyR..52c5502C}. Hence, we suggest that the steep velocity structure function can be caused by the expansion of the Sh~2-27 H{\scriptsize II} region or the dissipative range of incompressible turbulence.

\subsection{Magnetic field strength}\label{sec.B}
\begin{figure*}[!htbp]
\centering
\includegraphics[height = 0.45 \textwidth]{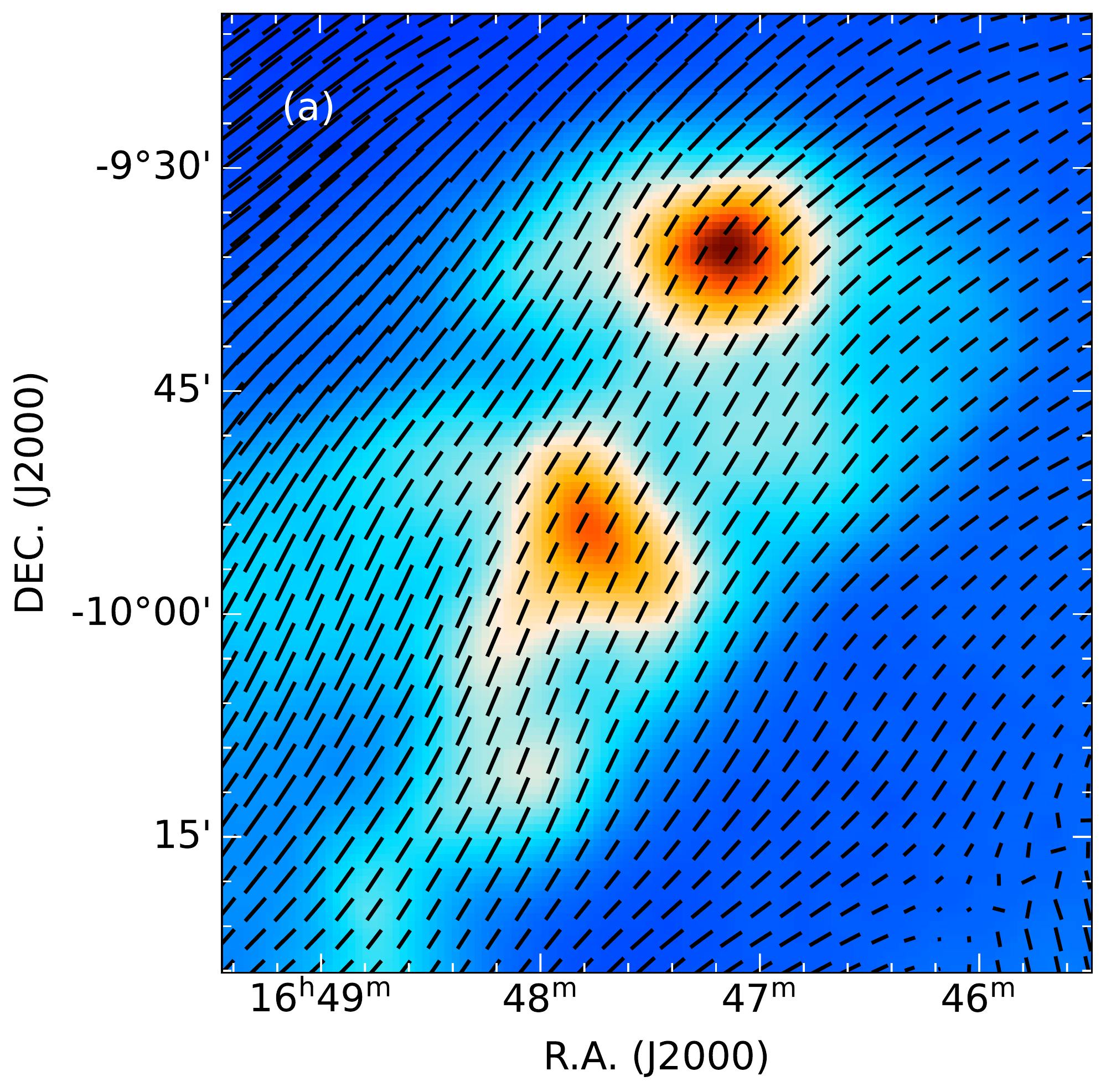}
\includegraphics[height = 0.45 \textwidth]{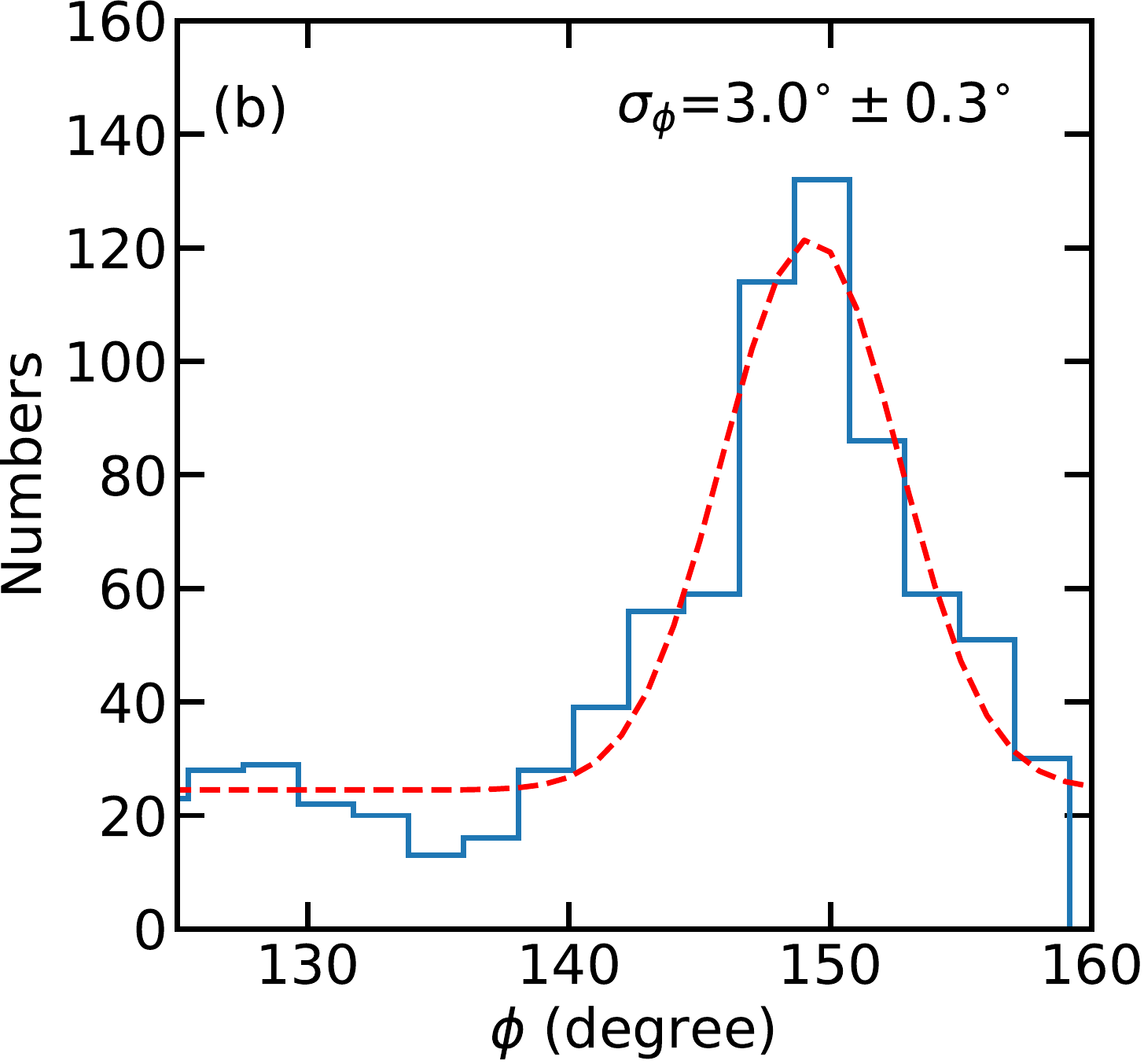}
\caption{{(a) Planck Stokes I continuum emission at 353 GHz is overlaid with magnetic field orientations. (b) Histogram distribution of the polarization angle fitted with a single Gaussian component.}\label{Fig:dcf}}
\end{figure*}

The large-scale magnetic field of Oph N1 has not been reported by previous Planck studies \citep[e.g.,][]{2019A&A...629A..96S}. We present the distribution of the plane-of-the-sky magnetic field toward Oph N1 in Fig.~\ref{Fig:dcf}a. The plane-of-the-sky magnetic field is nearly parallel to the elongation direction of Oph N1, which is different from the polarization configuration in OphN2 (L204) where the plane-of-the-sky magnetic field is perpendicular to its elongation direction \citep{1986ApJ...309..619M,1988ApJ...324..321H}. The large-scale magnetic field morphology is confirmed by the Planck dust polarization measurements (see Fig.~\ref{Fig:BB} in Appendix~\ref{app.bfield}). This is consistent with previous Planck results that the relative orientation is commonly observed to change progressively with increasing H$_{2}$ column densities, from mostly parallel to mostly perpendicular, which was interpreted a signature of Alfv{\'e}nic or sub-Alfv{\'e}nic turbulence \citep[e.g.,][]{2016A&A...586A.138P}.

As shown in Fig.~\ref{Fig:dcf}a, the dispersion of the polarization angles is small, indicating that its magnetic field plays an important role in this region. We adopt the Davis-Chandrasekhar-Fermi (DCF) method to estimate the magnetic field strength of the plane-of-sky component, $B_{\rm pos}$, in this region using the following formula \citep{1951PhRv...81..890D,1953ApJ...118..116C,2012ARA&A..50...29C}, 
\begin{equation}\label{f.b}
B_{\rm pos} = f\sqrt{4\pi \rho} \frac{\sigma_{\rm v}}{\sigma_{\phi}}  \;,
\end{equation}
where $f$ is the correction factor, $\rho$ is the cloud density, $\sigma_{\rm v}$ is the one-dimensional turbulence velocity dispersion, and $\sigma_{\phi}$ is the dispersion in polarization angle. 

Based on numerical simulations \citep{2001ApJ...546..980O,2001ApJ...561..800H}, $f$ is assumed to be 0.5 when $\sigma_{\phi}<$25\degree. Because C$^{18}$O ($J=1-0$) emission is widespread in the observed region, we use the critical density of C$^{18}$O ($J=1-0$) to represent the cloud density. Following the method introduced by \citet{2015PASP..127..299S}, we estimate the optical thin critical density of C$^{18}$O ($J=1-0$) to be 7$\times 10^{2}$~cm$^{-3}$ at a kinetic temperature of 10~K. On the other hand, we can roughly estimate the average cloud density by assuming a nearly prolate 3D shape, that is, the depth is close to its width ($\sim$1.1~pc) of the cometary cloud. This leads to an average cloud volume density of 7.4$\times 10^{2}$~cm$^{-2}$, which is in agreement with the optical thin critical density of C$^{18}$O ($J=1-0$). Hence, a cloud density of 7$\times 10^{2}$~cm$^{-3}$ is adopted in our estimate. For $\sigma_{\rm v}$, the median value of the non-thermal velocity dispersions is adopted (i.e., $\sigma_{\rm v}$=0.14$\pm$0.04~\kms). 
We estimate $\sigma_{\phi}$ from the histogram distribution of the polarization angle which is shown in Fig.~\ref{Fig:dcf}b. We perform a single-component Gaussian fitting on the histogram distribution, and this gives $\sigma_{\phi}$=3.0\degree$\pm$0.3\degree. Consequently, we obtain $B_{\rm pos}$=28$\pm$9~$\mu$G. The errors are derived with the Monte Carlo error analysis where 10,000 Monte Carlo simulations are carried out. The derived $B_{\rm pos}$ is slightly higher than other nearby molecular clouds' values which are calculated with the same method \citep[5--20~$\mu$G;][]{2016A&A...586A.138P}. This indicates that magnetic support might be  important for this cloud.

The DCF method is based on the assumption of the isotropic turbulent motions. However, our target may violate the assumption, which might overestimate $B_{\rm pos}$. Based on magnetohydrodynamic simulations,  \citet{2021A&A...647A.186S} propose an alternative relation to estimate $B_{\rm pos}$,
\begin{equation}\label{f.newb}
    B_{\rm pos} = \sqrt{2\pi \rho} \frac{\sigma_{\rm v}}{\sqrt{\sigma_{\phi}}}\,.
\end{equation}
\citet{2021A&A...647A.186S} compared the method with the simulations, and found a deviation of 17\%\,in $B_{\rm pos}$. 
Using Eq.~(\ref{f.newb}) and including the uncertainty of 17\%, we obtain $B_{\rm pos}$=9$\pm$3$\pm$2~$\mu$G. Although this method takes the anisotropic properties of turbulence into account, the estimated value from this method could be biased to lower values \citep{2021A&A...647A.186S}, because gravity cannot be neglected in our case (see discussions in Sect.~\ref{Sec:magcloud}).

Based on the two methods mentioned above, we can give a lower limit of $\sim$9~$\mu$G for $B_{\rm pos}$, which provides a lower limit for the total magnetic field strength, $B_{\rm t}$, that is, $B_{\rm t}\gtrsim$9~$\mu$G. In order to obtain the total magnetic field strength, one needs the magnetic field strength of the line-of-sight component, $B_{\rm los}$. \citet{1988ApJ...324..321H} measured the nearby cloud OphN2 (L204) with the HI Zeeman splitting, and found about 4.2~$\mu$G for the average $B_{\rm los}$. If the same $B_{\rm los}$ is assumed for Oph N1, we arrive at $B_{\rm t}\gtrsim$10~$\mu$G, which is comparable to the reported $B_{\rm t}$ of about 12~$\mu$G in OphN2 \citep{1988ApJ...324..321H}.
A study of polarized radio emission toward Sh 2-27 indicates the line-of-sight magnetic strengths of $-15$~$\mu$G and $+$30~$\mu$G in the near and far cloud
 \citep{2019MNRAS.487.4751T}. Because Oph N1 is located in front of Sh 2-27 as mentioned above, we assume $-15$~$\mu$G for the line-of-sight component toward Oph N1, which leads to $B_{\rm t}\gtrsim$18~$\mu$G. On the other hand, \citet{2004ApJ...600..279C} derived the statistical average relation, $B_{\rm pos}=\frac{\pi}{4} B_{\rm tot}$, which gives $B_{\rm t}\gtrsim$11~$\mu$G for our case. Hence, these different assumptions support $B_{\rm t}\gtrsim$10~$\mu$G for Oph N1.
 The $B_{\rm t}$ values give a three-dimensional Alfv{\'e}n velocity of $\gtrsim$0.5~\kms, where the three-dimensional Alfv{\'e}n velocity is defined as $\varv_{\rm A} = \frac{B_{\rm t}}{\sqrt{4\pi\rho}}$. The corresponding Alfv{\'e}n Mach number $\mathcal{M}_{\rm A}=\sqrt{3}\sigma_{\rm nt}/\varv_{\rm A}$ is about 0.5. Since the derived $\varv_{\rm A}$ is a lower limit, the corresponding $\mathcal{M}_{\rm A}$ is an upper limit. Hence, $\mathcal{M}_{\rm A}<$1 is robust. Therefore, we conclude that Oph N1 is globally sub-Alfv{\'e}nic.
 
\section{Discussion}\label{Sec:dis}
\subsection{Magnetically supported cloud}\label{Sec:magcloud}
We test cloud stability by comparing the observed properties with the magnetic critical condition which takes the projection effects into account  \citep[e.g.,][]{2014prpl.conf..101L}:
\begin{equation}\label{f.sta}
    B_{\rm t}\;(\mu {\rm G}) = 1.9\times 10^{-21} N_{\rm H,crit}\;({\rm cm}^{-2}) \;,
\end{equation}
where $N_{\rm H,crit}$ is the critical hydrogen column density. Adopting $B_{\rm t}$=10~$\mu$G, we obtain  $N_{\rm H,crit}$=5.3$\times 10^{21}$~cm$^{-2}$. The average H$_{2}$ column density is found to be 2.5$\times 10^{21}$~cm$^{-2}$, which is equivalent to $N_{\rm H}$=5$\times 10^{21}$~cm$^{-2}$ that is comparable to $N_{\rm H,crit}$. This indicates that magnetic field is at least comparable to gravity in Oph N1. Because $B_{\rm t}$=10~$\mu$G is a lower limit (see Sect.~\ref{sec.B}), the magnetic field should be even more important.

We also compare the magnetic pressure with thermal and turbulent pressures in order to estimate their relative roles in stabilizing the cloud. Magnetic pressure, $B_{\rm tot}^{2}/8\pi$, is estimated to be 4.0$\times 10^{-12}$ erg~cm$^{-3}$ when $B_{\rm tot}$ is set to be 10~$\mu$G. Adopting the C$^{18}$O ($J=1-0$) critical density of 7$\times 10^{2}$~cm$^{-2}$ and a kinetic temperature of 10~K (see discussions above), we derive thermal pressure to be 7$\times 10^{3}$~cm$^{-3}$~K (i.e., 9.7$\times 10^{-13}$ erg~cm$^{-3}$). Turbulent pressure is determined by $\frac{3}{2}\rho \sigma_{\rm nt}^{2}$, where $\rho$ is the density and $\sigma_{\rm nt}$ is the nonthermal velocity dispersion. Adopting a H$_{2}$ number density of 7$\times 10^{2}$~cm$^{-2}$ and $\sigma_{\rm nt}$=0.14~\kms\,(see Sect.~\ref{sec.kin}), we arrive at the turbulent pressure of 3.3$\times 10^{-12}$~erg~cm$^{-3}$. Because the adopted $B_{\rm tot}$ is a lower limit, this comparison suggests that magnetic pressure should be higher than thermal pressure and turbulent pressure. This result is also supported by our observed morphology that the cloud elongation is parallel to the plane-of-the-sky magnetic field, because such a configuration is expected in sub-Alfv{\'e}nic turbulence where the magnetic energy is above or comparable to the kinetic energy \citep{2013ApJ...774..128S}. Oph N1 is therefore supported against gravity mainly by the magnetic field.

\subsection{Energy dissipation}\label{sec.dissipation}
The widespread narrow line widths suggest large-scale subsonic turbulence in Oph N1, which tends to be more quiescent than most molecular clouds. A question can be raised here. How does the cloud reach the current dynamic state? Energy injection and dissipation should be the key to the question. 

Because Oph N1 lies at the edge of the Sh 2-27 H{\scriptsize II} region, the H{\scriptsize II} region may input kinetic energies into the clouds.  However, the Sh 2-27 H{\scriptsize II} region has a large projected diameter of $\sim$15~pc (see Fig.~\ref{Fig:overview}). Following the method used in \citet{2011A&A...527A..62B} and assuming an initial density of 1$\times 10^{3}$~cm$^{-3}$, we estimate the dynamic age of the Sh 2-27 H{\scriptsize II} region to be about 4~Myr, in agreement with the age ($\sim$3~Myr) of $\zeta$~Oph derived from evolutionary models~\citep{2011MNRAS.410..190T}. The numerical simulations of \citet{2013MNRAS.436..859M} suggest that the kinetic energy input from the Sh 2-27 region becomes negligible after the first 1.5~Myr (see their Fig.~9). This confirms that, at 4~Myr, Sh 2-27 is an old H{\scriptsize II} region that has already lost most of its kinetic energy. Oph N1 is about 4~pc away from $\zeta$~Oph, and the radiation from OB stars varies as the inverse square of distance, so the radiation should not play an important role. On the other hand, Oph N1 is also not active in star formation, so its internal feedback is also not important. Since the cloud is likely stabilized by magnetic fields (see Sect.~\ref{Sec:magcloud}), the energy input from gravity should be also negligible. These facts suggest that there are no considerable internal and external kinetic energy input toward Oph N1 recently. 

Energy dissipation by turbulent cascade might cause the low level of turbulence. We estimate the timescale of energy dissipation by turbulence cascade in order to evaluate its role. The timescale of energy dissipation by turbulent cascade is characterised by the crossing time , $\tau_{\rm c}$, that is determined by the size of the cloud, $L$, and velocity dispersion, $\sigma_{\rm nt}$, (i.e., $\tau_{\rm c} = \frac{L}{\sigma_{\rm nt}}$). In our case, we adopt the width of 1.1~pc and the median non-thermal velocity dispersion of 0.14~\kms, which results in about 8~Myr for the crossing time. Because this process can cause lower $\sigma_{\rm nt}$ with time, $\sigma_{\rm nt}$ should be higher in earlier stages, which implies that the timescale of about 8~Myr can be overestimated.  

Alternatively, the energy dissipation by the ion-neutral friction can be potentially more important, because the cloud is globally sub-Alfv{\'e}nic. We can estimate the timescale for energy dissipation by the ion-neutral friction, $\tau_{\rm diss,amb}$, with Eq.~(13) in \citet{2013A&A...560A..68H}:

\begin{equation}\label{f.amb}
\tau_{\rm diss,amb} = \frac{2\gamma_{\rm damp}\rho_{\rm i}}{\varv_{\rm A}(2\pi/\lambda)^{2}}\;,
\end{equation}
We adopt the damping rate, $\gamma_{\rm damp}$, to be $3.5\times 10^{13}$~cm$^{3}$~g$^{-1}$~s$^{-1}$ and the ion density, $\rho_{\rm i}$, to be $\rho_{\rm i}=C\sqrt{\rho_{\rm n}}$ \citep{1979ApJ...232..729E,2013A&A...560A..68H}, where $C=3\times 10^{-16}$~cm$^{-3/2}$~g$^{1/2}$ and the neutral density, $\rho_{\rm n}$, is assumed to be the C$^{18}$O ($J=1-0$) critical density of 7$\times 10^{2}$~cm$^{-2}$. The Alfv{\'e}nic speed is set to be 0.5~\kms\,(see Sect.~\ref{sec.B}), and the wavelength, $\lambda$, is assumed to be equal to the width ($\sim$1.1~pc) of Oph N1. This gives the timescale of 2.7~Myr. If we take the Alfv{\'e}nic speed of $>$0.5~\kms\,and the wavelength of $<$1.1~pc, the timescale can become significantly shorter.   

If we believe that the structure is formed after the interaction with the H{\scriptsize II} region, the age of H{\scriptsize II} region places an upper limit on the formation of the structure (i.e., $\lesssim$3~Myr). The age is much lower than the timescale of energy dissipation by turbulent cascade, but is comparable to or higher than the timescale of energy dissipation by ion-neutral friction. As discussed in Sect.~\ref{sec.vsf}, the classic turbulent cascade cannot solely explain the observed velocity structure function, which is indicative of additional dissipation mechanisms. Therefore, we suggest that the energy dissipation by the ion-neutral friction should play an important role in forming the observed large-scale subsonic turbulence. 




\section{Summary}\label{Sec:sum}
We have simultaneously mapped Ophiuchus North 1 (Oph N1) in CO ($J=1-0$) and C$^{18}$O ($J=1-0$) with the PMO-13.7 m telescope to study its physical properties. Our main findings are summarized as follows:
\begin{itemize}

\item[1.] 
We find that most of the whole C$^{18}$O emitting regions have Mach numbers of $\lesssim$1, demonstrating the extended subsonic turbulence up to a scale of $\gtrsim$1.5~pc. Based on the measurements of the local velocity gradients, the velocity field indicates the presence of velocity shear, while the contributions of turbulence on the local velocity gradients should not be neglected.

\item[2.] Oph N1 exhibits a head-tail morphology at the edge of the Sh 2-27 H{\scriptsize II} region. The excitation temperatures are within the range of 7.5--12.0~K with a median value of 8.4~K. High excitation molecular gas is found in the outer regions, which might be caused by external heating. The C$^{18}$O fractional abundances with respect to H$_{2}$ are within the range of (0.2--1.7)$\times 10^{-7}$. We show the presence of regions of C$^{18}$O depletion towards the center of the two Planck cold clumps.

\item[3.] The plane-of-the-sky magnetic field is nearly parallel to the long axis of Oph N1. Based on the polarization measurements at 353 GHz, we estimate the magnetic field strength of the plane-of-sky component to be $\gtrsim$9~$\mu$G, and the total magnetic field strength should be $\gtrsim$10~$\mu$G. We find that Oph N1 is globally sub-Alfv{\'e}nic, and is supported against gravity mainly by the magnetic field.

\item[4.] We construct the second-order velocity structure function of Oph N1 from the C$^{18}$O ($J=1-0$) velocity centroids. The power-law index is found to be 1.30$\pm$0.03 in the spatial range of 0.03--0.5~pc. We suggest that the steep velocity structure function can be caused by the expansion of the Sh~2-27 H{\scriptsize II} region or the dissipative range of incompressible turbulence. Comparing the different timescales of dissipation, we suggest that the energy dissipation by ion-neutral friction should play an important role in forming the observed widespread subsonic turbulence.


\end{itemize}



\section*{ACKNOWLEDGMENTS}\label{sec.ack}
We acknowledge the PMO-13.7 m staff for their assistance with our observations. This work was partially supported by the National Key R\&D Program of China under grant 2017YFA0402702. W.S.Z. is supported by NSFC grant 12173102. GXL acknowledges supports from NSFC grant W820301904 and 12033005. This research has made use of NASA's Astrophysics Data System. This work also made use of Python libraries including Astropy\footnote{\url{https://www.astropy.org/}} \citep{2013A&A...558A..33A}, NumPy\footnote{\url{https://www.numpy.org/}} \citep{5725236}, SciPy\footnote{\url{https://www.scipy.org/}} \citep{jones2001scipy}, Matplotlib\footnote{\url{https://matplotlib.org/}} \citep{Hunter:2007}, APLpy \citep{2012ascl.soft08017R}, GaussPy+ \footnote{\url{https://github.com/mriener/gausspyplus}} \citep{2019A&A...628A..78R}, and magnetar\footnote{\url{https://github.com/solerjuan/magnetar}} \citep{Soler2013}. This publication makes use of data products from the Wide-field Infrared Survey Explorer, which is a joint project of the University of California, Los Angeles, and the Jet Propulsion Laboratory/California Institute of Technology, funded by the National Aeronautics and Space Administration. We would like to thank the anonymous referee for a careful review of this article and valuable comments. 

\bibliographystyle{aa}
\bibliography{references}

\clearpage
\begin{appendix}


\section{Decomposition of CO ($J=1-0$) spectra}\label{app.b}
\begin{figure*}[!htbp]
\centering
\includegraphics[height = 0.35 \textwidth]{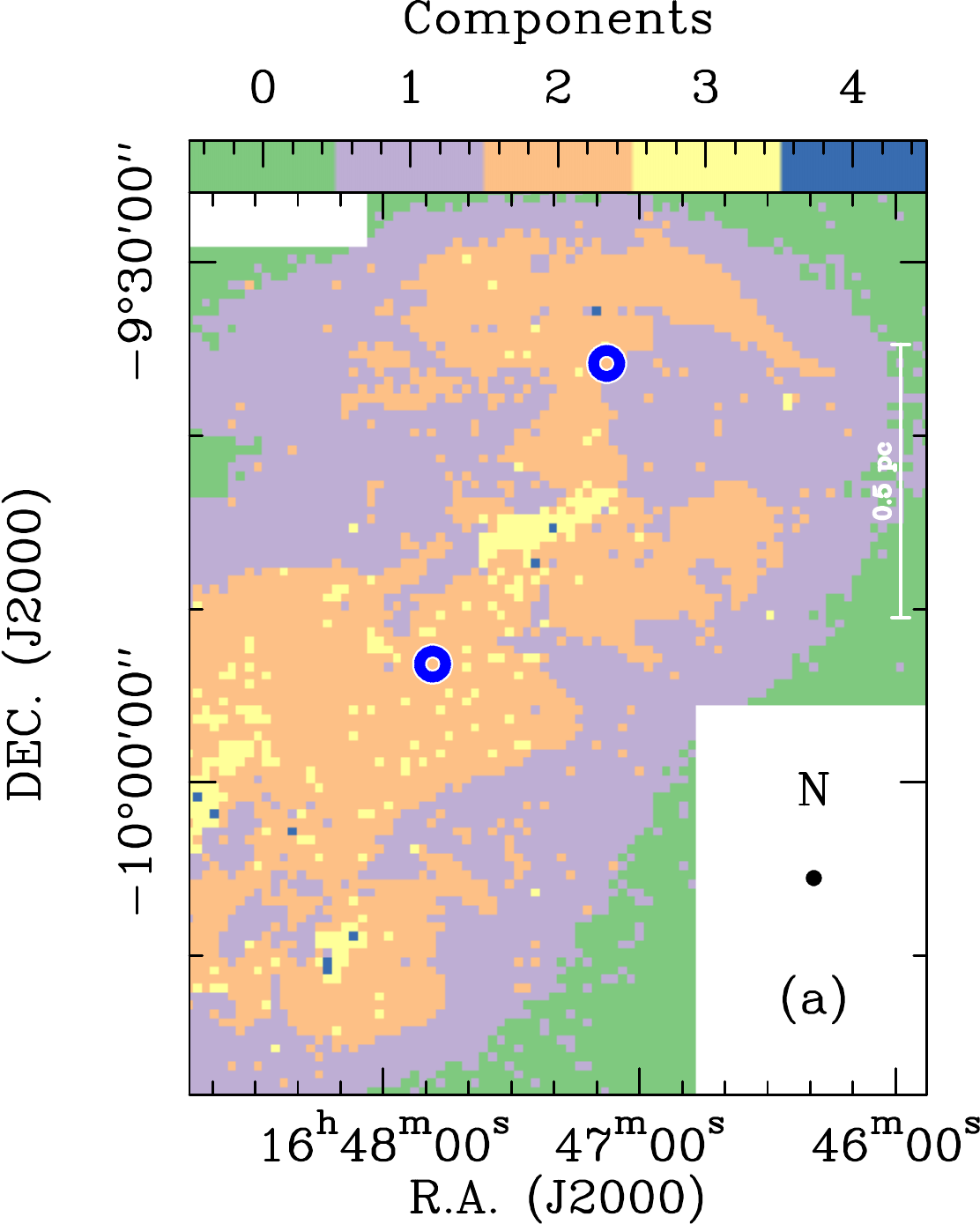}
\includegraphics[height = 0.35 \textwidth]{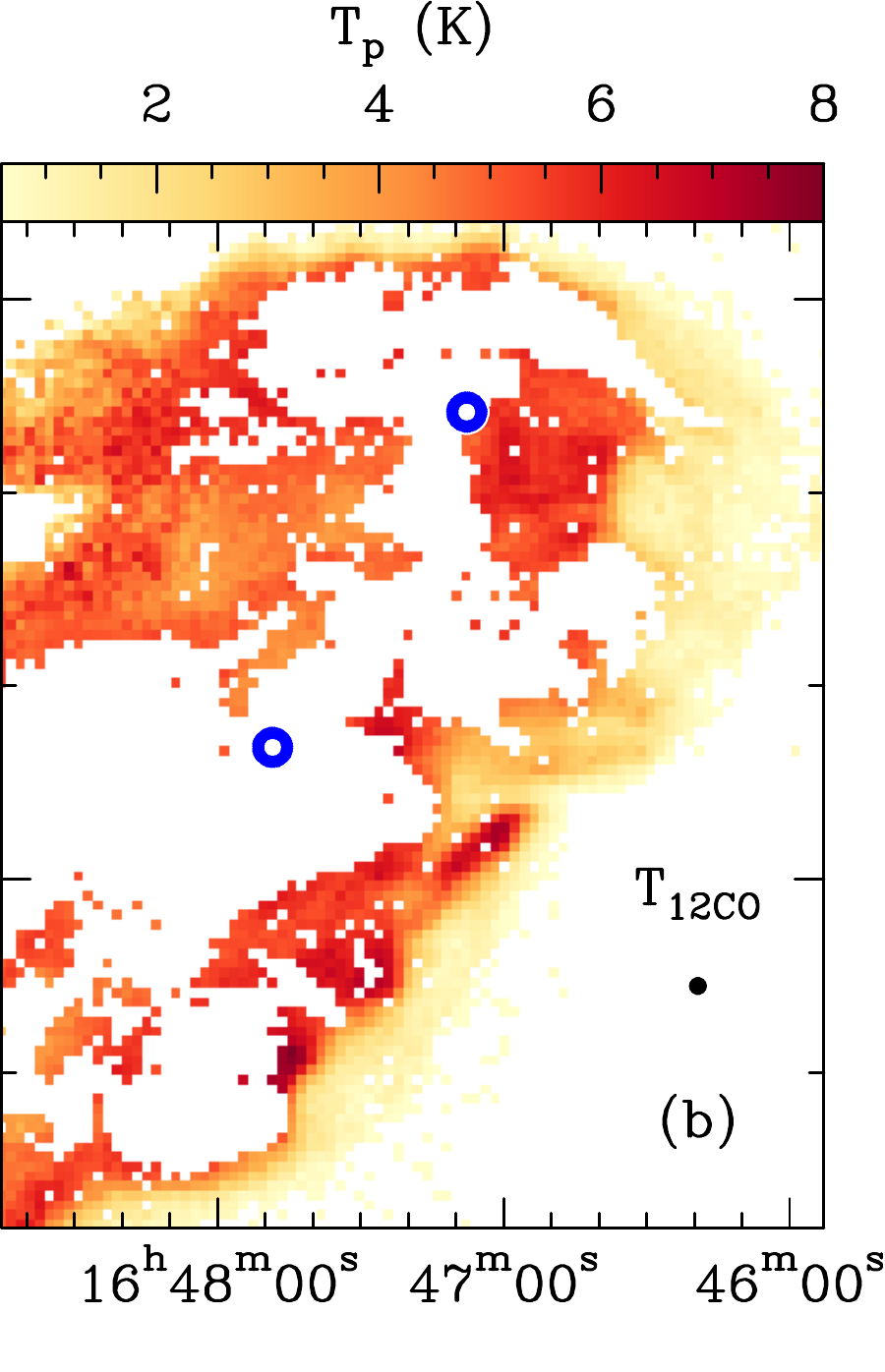}
\includegraphics[height = 0.35 \textwidth]{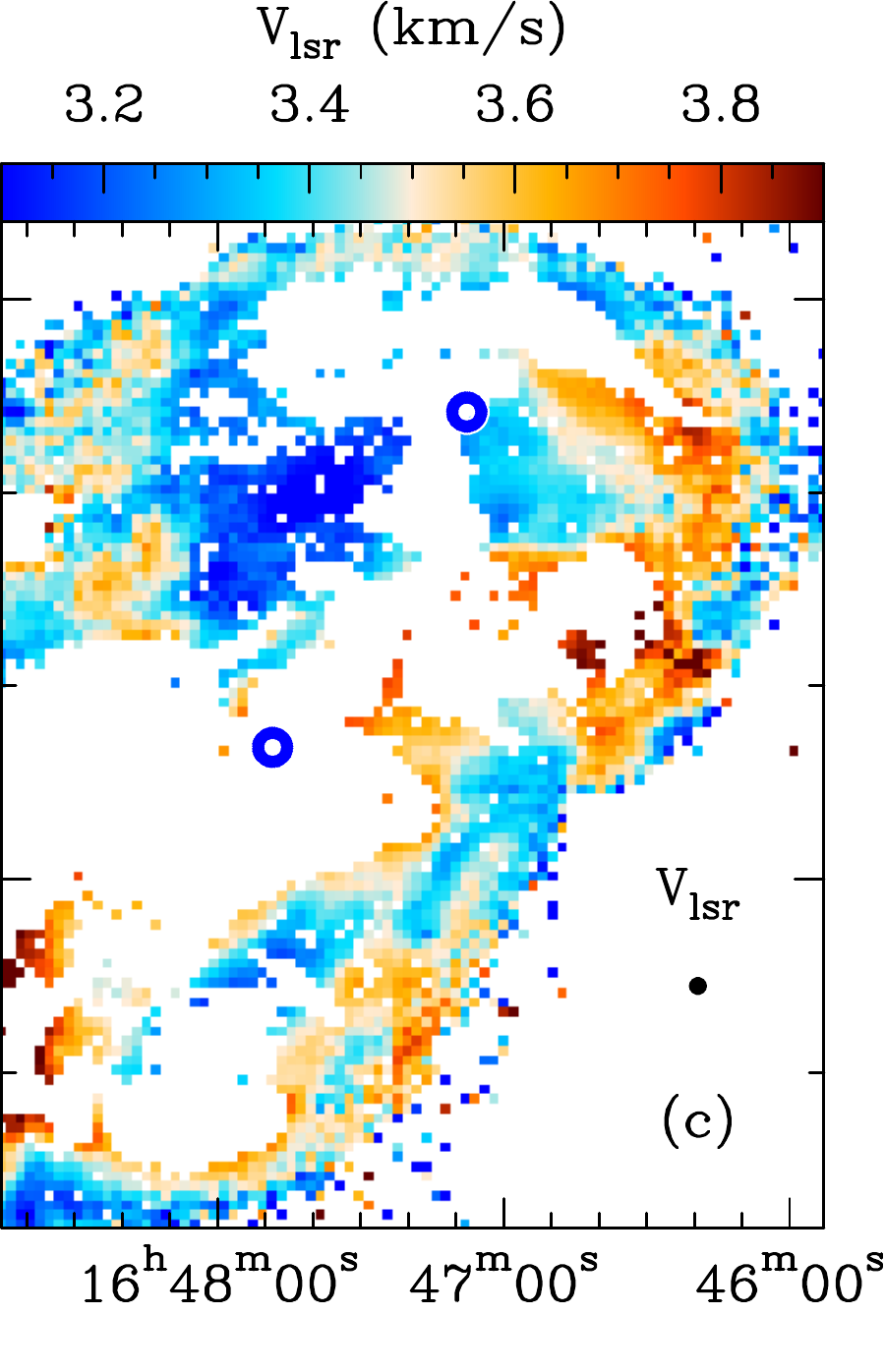}
\includegraphics[height = 0.35 \textwidth]{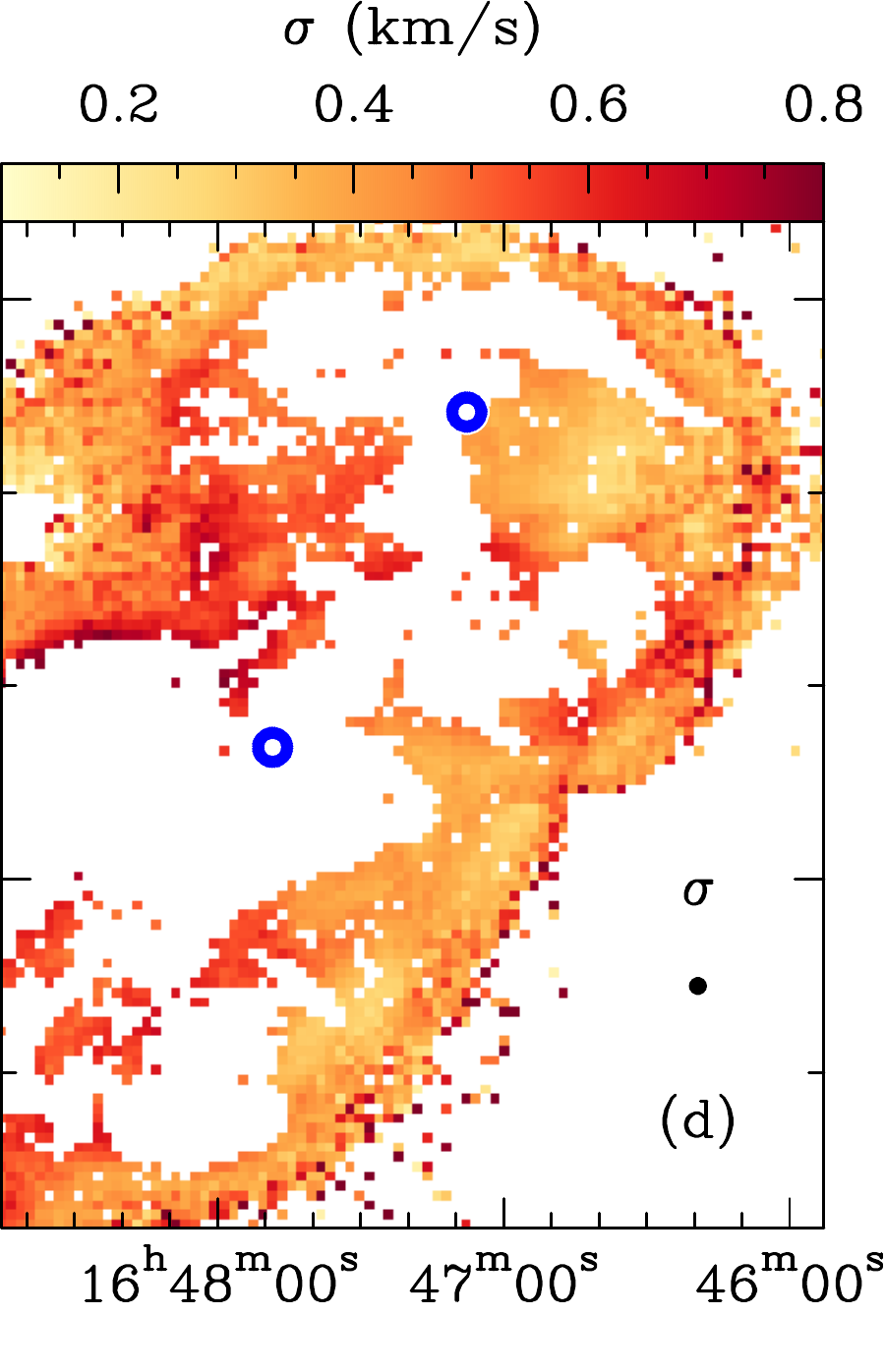}
\caption{Decomposition of CO ($J=1-0$) data. (a) Distribution of the number of the fitted velocity components. (b) Fitted peak intensities for the pixels showing a single velocity component. (c) Fitted velocity centroids for the pixels showing a single velocity component. (d) Fitted velocity dispersions for the pixels showing a single velocity component. In each panel, the beam size is shown in the lower right corner, the two Planck cold clumps (G008.67+22.14 and G008.52+21.84) are indicated by the two open blue circles.
\label{Fig:decom12}}
\end{figure*}

We fit the Gaussian profiles to the CO ($J=1-0$) data cube, using the fully automated Gaussian decomposition package, GAUSSPY+ \citep{2019A&A...628A..78R} that is based on the GAUSSPY algorithm \citep{2015AJ....149..138L}. We successfully obtained the fitting results toward 6395 spectra, and the results are presented in Fig.~\ref{Fig:decom12}. In Fig.~\ref{Fig:decom12}a, We find that 3521 spectra show one single velocity component and the rest 2874 display multiple velocity components. Because we cannot disentangle whether the multiple velocity components are caused by self-absorption (see Sect.~\ref{sec.mor}) or multiple gas structures, we only investigate the spectral results showing a single velocity component, and the fitted results are shown in Figs.~\ref{Fig:decom12}b--\ref{Fig:decom12}d.   
In Fig.~\ref{Fig:decom12}d, we find 1106 pixels with velocity dispersions of $<$0.38~\kms, suggesting that $\mathcal{M}<$2 even without taking opacity broadening and local velocity gradients into account. Therefore, the results further support the low level of turbulence across Oph N1. 

\section{Line-ratio map}\label{app.ratio}
\begin{figure}[!htbp]
\centering
\includegraphics[width = 0.45 \textwidth]{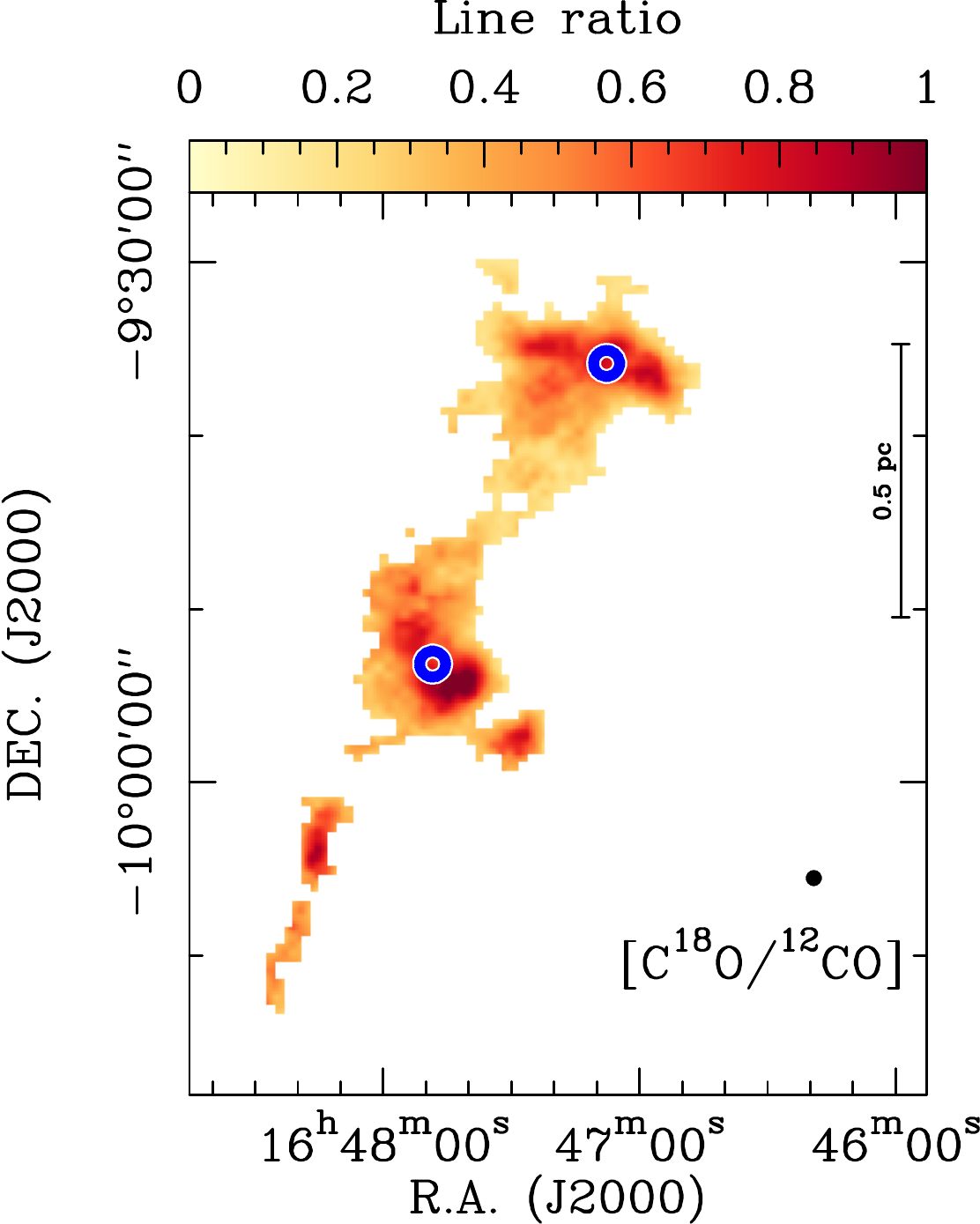}
\caption{Distribution of the line ratios between C$^{18}$O and CO at the peak velocity of C$^{18}$O. \label{Fig:ratio}}
\end{figure}

We first derived the peak intensity map of C$^{18}$O ($J=1-0$) that is clipped at 3$\sigma$. The corresponding peak intensity map of CO ($J=1-0$) is obtained at the peak velocity of C$^{18}$O ($J=1-0$). The line-ratio map is directly estimated by the ratio of the two maps, and Fig.~\ref{Fig:ratio} presents the distribution of the line ratios between C$^{18}$O and CO. The line ratios of $>$1 are found toward G008.52+21.84, which is caused by the significant CO self-absorption (see Fig.~\ref{Fig:spec}c). There are also high line ratios of $>$0.5 toward G008.67+22.14, which can also be caused by CO self-absorption (see Fig.~\ref{Fig:spec}a). This indicates widespread $^{12}$CO (1--0) self-absorption in observed regions (see also Sect.~\ref{sec.mor}).

\section{Large-scale magnetic field morphology toward Ophiuchus North}\label{app.bfield}
\begin{figure*}[!htbp]
\centering
\includegraphics[width = 0.95 \textwidth]{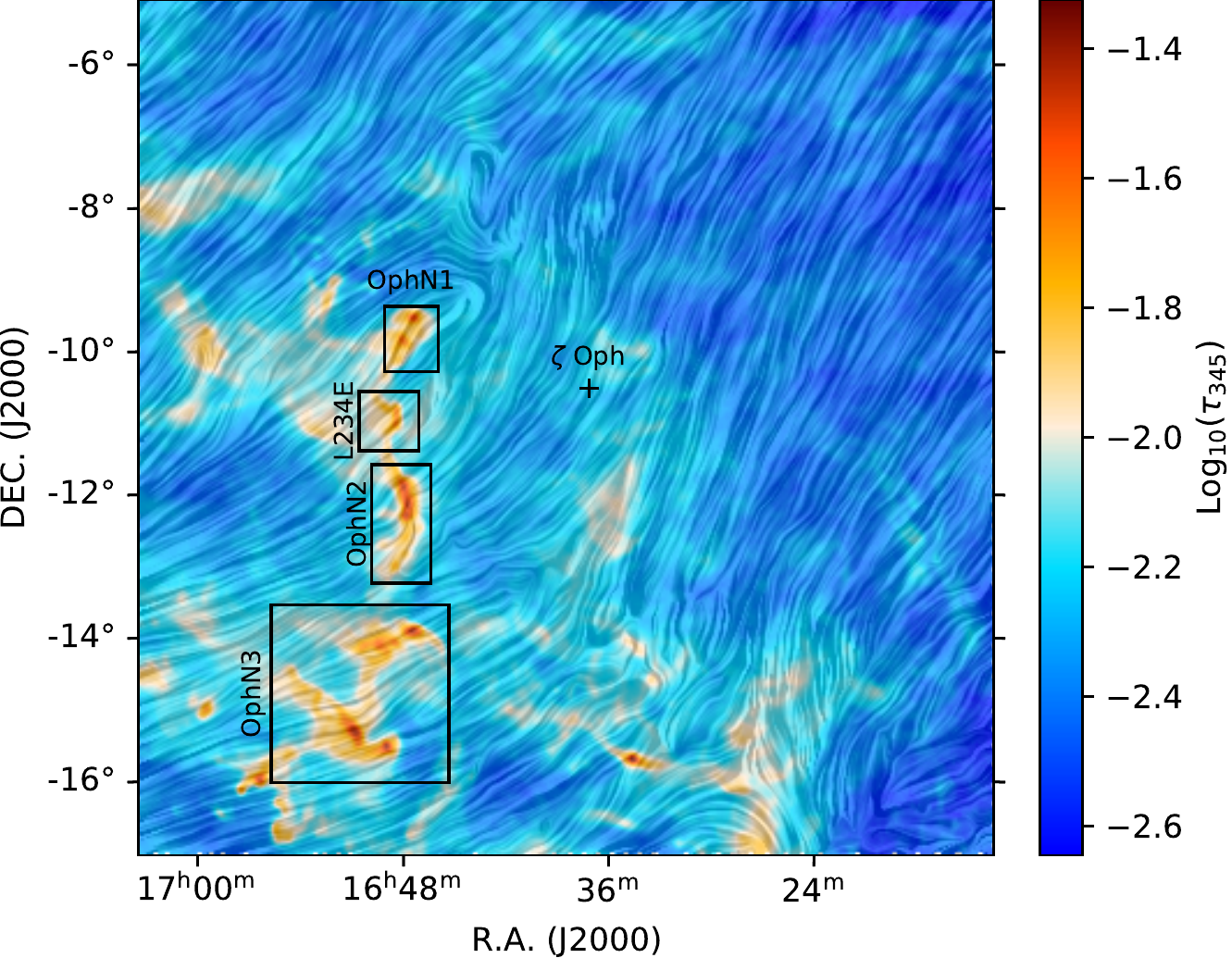}
\caption{Plane-of-the-sky magnetic field and $\tau_{345}$ measured by Planck toward Ophiuchus North. The overlaid pattern, produced using the line integral convolution (LIC) method \citep{Cabral93}, indicates the orientation of magnetic field lines. The marked regions are the same as in Fig.~\ref{Fig:overview}. \label{Fig:BB}}
\end{figure*}

Figure~\ref{Fig:BB} presents the large-scale magnetic field morphology of Ophiuchus North that is traced by the Planck dust polarization data. This figure shows that the plane-of-the-sky magnetic field is nearly parallel to
the elongation direction of Oph N1 and becomes perpendicular to the elongation directions of L234E and OphN2.

\end{appendix}

\end{document}